\documentclass[prx,twocolumn,superscriptaddress,aps,10pt, longbibliography=false]{revtex4-2}
\usepackage{amsthm}
\usepackage{amsmath}
\usepackage{amssymb}
\usepackage{amsfonts}
\usepackage[latin1,utf8]{inputenc}
\usepackage[american]{babel}
\usepackage{graphicx,xcolor,bbold,titlesec}

\usepackage{MnSymbol}
\usepackage{braket}
\usepackage{bm}
\usepackage{verbatim}

\usepackage[colorlinks,
citecolor=teal,
linkcolor=teal,
urlcolor=teal]{hyperref}
\usepackage{tikz,ifthen}
\usepackage{bbold}
\usepackage{orcidlink}
\usepackage{tikz-network}
\usetikzlibrary{patterns,decorations.pathreplacing,calligraphy}
\usetikzlibrary{shapes,arrows.meta,decorations.pathmorphing}

\definecolor{myyellow}{RGB}{78, 154, 6}
\definecolor{myorange}{RGB}{204, 85, 0}
\definecolor{myorangel}{RGB}{255,204,153}
\definecolor{myblue}{RGB}{52, 101, 164}
\definecolor{Armangreen}{RGB}{0,150,0}
\newcommand{\kket}[1]{|#1\rrangle}
\newcommand{\bbra}[1]{\llangle #1|}
\newcommand{\bbrakket}[2]{\llangle #1|#2\rrangle}

\newcommand{\Wg}{{\text{Wg}}}
\newcommand{\Ex}{\mathbb{E}}
\newcommand{\tr}{\text{tr}}
\newcommand{\G}{\text{G}}

\usepackage{xcolor}

\newtheorem*{conjecture}{Conjecture}

\newcommand{\titleinfo}{Universality in the Anticoncentration of Chaotic Quantum Circuits}

\begin{document}

\title{\titleinfo}

\author{Arman Sauliere~\orcidlink{0009-0004-4065-0320}}
\thanks{These two authors contributed equally.}
\affiliation{Laboratoire de Physique Th\'eorique et Mod\'elisation, CNRS UMR 8089, CY Cergy Paris Universit\'e, 95302 Cergy-Pontoise Cedex, France}

\author{Beatrice Magni~\orcidlink{0009-0009-8577-0525}}
\thanks{These two authors contributed equally.}
\affiliation{Institute f\"{u}r Theoretische Physik, Universit\"{a}t zu K\"{o}ln, Z\"{u}lpicher Straße 77, D-50937 K\"{o}ln, Germany}

\author{Guglielmo Lami~\orcidlink{0000-0002-1778-7263}}
\affiliation{Laboratoire de Physique Th\'eorique et Mod\'elisation, CNRS UMR 8089, CY Cergy Paris Universit\'e, 95302 Cergy-Pontoise Cedex, France}

\author{Xhek Turkeshi~\orcidlink{0000-0003-1093-3771}}
\affiliation{Institute f\"{u}r Theoretische Physik, Universit\"{a}t zu K\"{o}ln, Z\"{u}lpicher Straße 77, D-50937 K\"{o}ln, Germany}

\author{Jacopo De Nardis~\orcidlink{0000-0001-7877-0329}}
\affiliation{Laboratoire de Physique Th\'eorique et Mod\'elisation, CNRS UMR 8089, CY Cergy Paris Universit\'e, 95302 Cergy-Pontoise Cedex, France}

\begin{abstract}
We identify a \emph{universal functional form} that governs anticoncentration in random quantum circuits—one that holds across diverse circuit architectures and depths, and crucially remains valid even at finite system sizes and shallow depth. 
We support this claim through analytical results for ensembles of random tensor-network states and random-phase models. 
This compact, universal expression for the output bitstring probability distribution is fully characterized by just two fitting parameters, as validated through extensive numerical simulations.  
Our findings underscore the pivotal role of finite-size and finite-depth effects in shaping anticoncentration and introduce a practical framework for benchmarking quantum devices using shallow circuits, thereby enabling validation of systems significantly larger than previously accessible.
\end{abstract}

\maketitle

Recent progress in quantum platforms has dramatically expanded our ability to prepare, control, and measure many-body quantum states, offering unprecedented opportunities to explore the principles of quantum matter. 
As a consequence, quantum circuits--once viewed primarily as algorithmic constructs--have emerged as a crucial conceptual tool. They furnish a flexible framework to describe a wide range of quantum phenomena, bridging diverse fields such as quantum chaos, thermalization, black-hole physics, and computational complexity~\cite{dalzell2022random,eisert2023computational,Watrous_2018,Morvan2023,2023Google,Fisher2023}. 

A key property in the study of these systems is \emph{anticoncentration}~\cite{bertoni2024shallow,bertini2020scrambling,cioli2024approximateinversemeasurementchannel,mark2023benchmarking,mark2024federica,fefferman2024anticoncentration,magni2025anticoncentrationcliffordcircuitsbeyond,christopoulos2024universaldistributionsoverlapsunitary}, which captures the extent to which an ensemble of quantum states spreads over the computational basis. 
From the perspective of randomness, an anticoncentrated ensemble has overlaps that are approximately Porter-Thomas distributed, mirroring the predictions of random matrix theory. 
In strongly chaotic systems, it is expected that such universal behavior emerges in logarithmic depth~\cite{dalzell2022random}, a phenomenon grounded in both exact calculations on tractable random-circuit ensembles and extensive numerical verification~\cite{christopoulos2024universaldistributionsoverlapsunitary,Lami2025,Tirrito2024}.

Nevertheless, the path toward complete Porter--Thomas behavior can exhibit finite-size corrections and nontrivial scaling, prompting the fundamental question: \emph{To what extent do these corrections depend on the microscopic details of the circuit architecture?} In this work, we provide a comprehensive analysis of the approach to anticoncentration, showing that large classes of chaotic quantum circuits share a \emph{universal} crossover characterized by just a few simple parameters.

\begin{figure}[t!]
     \centering
     \includegraphics[width=0.92\linewidth]{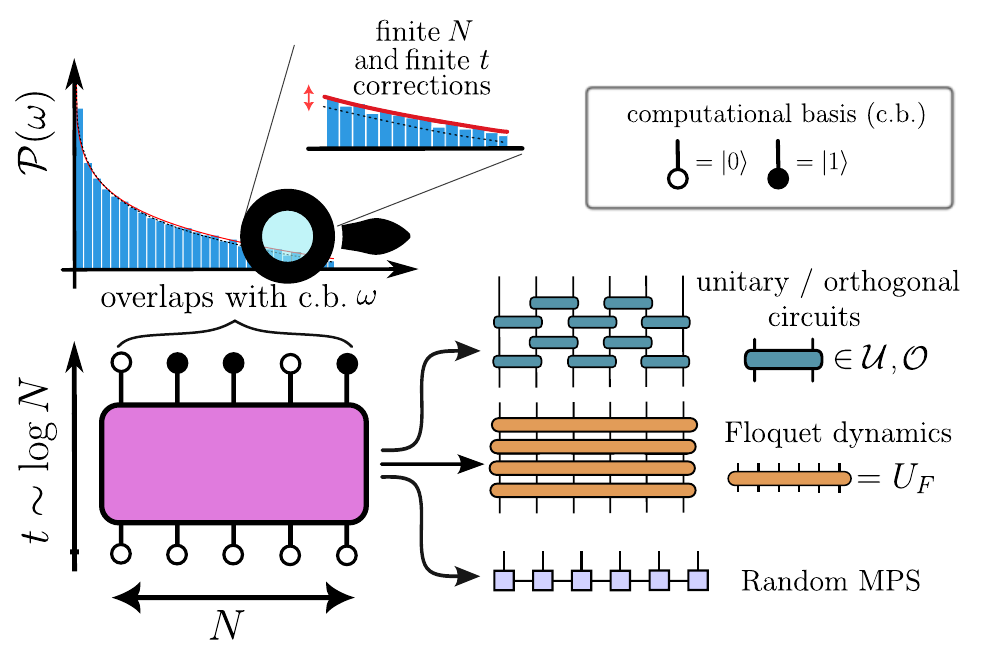}
     \caption{  Illustrative sketch of the work. We examine various types of quantum states, including random Matrix Product States (MPS), outputs of random brickwork quantum circuits and Floquet dynamics at time $t$. We study the distribution $\mathcal{P}(\omega)$ of their overlaps $\omega$ with the computational basis (c.b.). We show that in the regime $t \lesssim \log N$, where $N$ is the number of qubits, all these models exhibit a universal form for the finite-$N$ or finite-depth corrections to $\mathcal{P}(\omega)$.}
     \label{fig:sketch}
 \end{figure}

To anchor these ideas, we begin by analyzing ensembles of random tensor network states~\cite{hayden2016holographic,qi2017holographic,cheng2024random,piroli2020a,schollwock2011the,biamonte2020lectures,silvi2019the,Orus_2014,Ranabhat_2022,Ranabhat_2024,Verstraete_2004,Jordan_2008,PhysRevLett.124.037201,Corboz_2011}, where the disorder averaging enables exact, closed-form expressions for the inverse participation ratio and related measures of delocalization. 
We find that both the leading scaling behavior and the dominant finite-size corrections collapse onto a universal curve when plotted as a function of a single dimensionless parameter, $ N/w^t$, with $N$ the system size, $t$ the time or circuit depth, and $w$ encoding circuit-specific details. 
These analytical results are further supported by numerical simulations of Haar-random unitary circuits and chaotic Floquet circuits, all of which collapse onto the same scaling form. 

Our paper is organized as follows. We begin by summarizing the key aspects of Weingarten calculus and tensor networks that are relevant to our derivations. Next, we analytically derive the universal form of the overlap distribution, including subleading finite-size terms, for random tensor network states. We corroborate the predictions through large-scale simulations of unitary and orthogonal brickwork quantum circuits and discuss how these findings naturally extend to generic, chaotic quantum evolutions. Finally, we demonstrate that our findings can be used to successfully benchmark the output of large quantum circuits, even with relatively low depths.

\section{Anticoncentration in quantum systems}
To provide background for the following discussion, we briefly review the concept of anticoncentration and its quantification in the context of many-body systems. 
Consider a system of $N$ qudits, each with a local Hilbert space dimension $d$. 
We denote $D=d^N$ the total and we define the computational basis as $\mathcal{B} = \{\ket{\pmb{x}}\}_{\pmb{x}=0}^{D-1}$. 
Given an ensemble of pure state $\mathcal{D}=\{\ket{\psi}\}$, anticoncentration, tied to the notion of Hilbert space delocalization~\cite{luitz2014universal,sierant2022universal,mace2019multifractal,turkeshi2024hilbert,claeys2024fockspace}, quantifies the extent to which an ensemble of many-body wave functions spreads over the computational basis, providing a measure of scrambling in a quantum system. 

In this context, anticoncentration characterizes the statistical properties of overlaps $p_{\pmb{x}} \equiv |\langle \pmb{x}|\psi \rangle|^2$. 
A powerful proxy for assessing anticoncentration is given by the inverse participation ratios and the associated participation entropies, both defined with respect to the computational basis, respectively, as
\begin{equation}
    I_{k}(\ket{\psi}) \equiv \sum_{\pmb{x}} |\braket{\pmb{x}|\psi}|^{2k}\;, \; S_k\equiv \frac{1}{1-k}\ln[I_k]\;.\label{eq:defiksk}
\end{equation}

We note that $I_1=1$ corresponds to the normalization condition, and $k=2$ is referred to in the literature by collision probability~\cite{dalzell2022random,braccia_computing_2024}. 
A state is fully localized when $I_k=1$ for any $k$, leading to $S_k=0$. 
Similarly, we say a state is localized when $I_k\simeq S_k \simeq O(1)$ is independent of system size. 
Nevertheless, most states in a many-body Hilbert space are spread through the whole computational basis, and typically $S_k= D_k N + c_k$, with $D_k$ known as the multifractal dimension~\cite{luitz2014participation}.

Our focus will be on the average inverse participation entropy over the distribution of states $\mathcal{D}$, defined by
\begin{equation}
    {I}_k^{\mathcal{D}} = \mathbb{E}_{\psi \sim \mathcal{D}} [I_k(\ket{\psi})] = D \mathbb{E}_{\pmb{x} \sim \mathcal{B}, \psi \sim \mathcal{D}} [|\braket{\pmb{x}|\psi}|^{2k}],\label{eq:mustard}
\end{equation}
where $\mathbb{E}_{\psi\sim\mathcal{D}} [\dots] $ is the expected value with respect to the distribution $\mathcal{D}$. 
When the ensemble is local unitary invariant, the IPRs correspond up to a multiplicative constant to the moments of the random variable $\omega=D|\braket{\pmb{0}|\psi}|^2$, which represents the overlap of the states in $\mathcal{D}$ with the computational basis state $\ket{\pmb{0}}$. 
Specifically, the $k-$th moment of $\omega$ is given by $\Ex[\omega^k]=D^{k-1}I_k^{\mathcal{D}}$. 

Knowledge of all the moments is equivalent to knowing the \textit{full bitstring probability of probability distribution} of $\omega$, which is defined in general by 
\begin{equation}
    \mathcal{P}(\omega) \equiv \mathbb{E}_{\pmb{x} \sim \mathcal{B}, \psi \sim \mathcal{D}} \left[ \delta \left( \omega - D |\braket{\pmb{x}|\psi}|^2 \right) \right] \, .
\end{equation}
 Within this framework, a distribution of states is said to be fully anticoncentrated if $\mathcal{P}(\omega)$ closely approximates the corresponding distribution obtained when $\mathcal{D}$ is drawn from the Haar ensemble, see Sec.~\ref{subsec:AnticoncentrationHaar}.

The anticoncentration properties of many-body systems garnered significant attention in recent years, as they are directly related to the ability of the quantum circuit dynamics to span over all the accessible Hilbert space and achieve deep thermalization, cf. Ref.~\cite{gross2007evenly,mele2024introduction,brandao2016local,ippoliti2022solvable,fava2024designsfreeprobability,cotler2023emergent,claeys2022emergentquantum,pappalardi2022eigenstate,fritzsch2024microcanonicalfreecumulantslattice,foini2024outofequilibriumeigenstatethermalizationhypothesis,pappalardi2024eigenstatethermalizationfreecumulants,Lami_2024,Lami2025}. 
In this work, we establish that, irrespective of the specific setup—provided the dynamics are chaotic—the distribution of overlaps follows a \textit{universal form whose validity extends well beyond the large-time and large-system-size limits}, capturing also subleading and even sub-subleading corrections. This universality enables the use of anticoncentration as a powerful tool to benchmark quantum circuit outputs by probing only shallow depths, even for systems of very large size. As a result, our approach opens the door to scalable benchmarking protocols and classical simulations for regimes previously considered intractable.

\section{Methods}
\label{sec:methods}
Our work combines analytical arguments with exact numerical simulations obtained through tensor network~\cite{schollwock2011the,Orus_2014} and replica tensor network methods~\cite{nahum2017quantum,PhysRevX.8.031058,turkeshi2024quantummpembaeffectrandom,turkeshi2024magicspreadingrandomquantum,braccia_computing_2024}. 
This section provides an overview of the key techniques used, including the graphical formalism employed to compute tensor contractions.

\subsection{Weingarten calculus} \label{subsec:wmethodsa}
We start by reviewing the \textit{Weingarten calculus} \cite{CHOI1975285,mele2024introduction},  presented in the vectorization formalism.
In this approach, all operators $A$ are reshaped as vectors $\kket{A}$ such that their inner product is given by $\bbrakket{A}{B} =\tr(A^\dagger B)$ and the action of conjugation by a unitary $E$ is expressed as $\kket{E A E^\dagger} =(E\otimes E^\ast)\kket{A}$~\cite{mele2024introduction}. 
Our interest lies in the computation of the $k$-moments of Haar-distributed gates acting over a Hilbert space of dimension $q$ on finite-depth circuits
\begin{equation}
    \mathbf{E}_t = \prod_{s=1}^t\left(\prod_{\lambda\in \Lambda_s}E_\lambda\right).\label{eq:circuit}
\end{equation}
In the above expression, $\lambda$ indicates the sites, out of the total $N$, on which the unitary gate $E$ acts, while  $\Lambda_s$ determines the active sites on a given time step, or circuit depth, $s$. 

As discussed below, Eq.~\eqref{eq:circuit} encompasses both brickwork random circuits built of nearest-neighboring gates, and staircase circuits on $r+1$ qudits defining random matrix product states (RMPS). 
A straightforward algebraic manipulation shows that computing the inverse participation ratios in Eq.~\eqref{eq:mustard} requires evaluating the expectation value of $k$-copies of the state
\begin{align}
I^\mathcal{D}_k&= D\Ex_{E_\lambda \sim \mathcal{E}}[\llangle{0,0}|^{\otimes k}(\mathbf{E}_t\otimes \mathbf{E}_t^*)^{\otimes k} \kket{\rho_0}^{\otimes k}] \nonumber \\ &= D\llangle{0,0}|^{\otimes k} \Ex_{E_\lambda \sim \mathcal{E}}[(\mathbf{E}_t\otimes \mathbf{E}_t^*)^{\otimes k}]\kket{\rho_0}^{\otimes k}.
\label{eq:IPR_circuit}
\end{align}
In the above expression, $\kket{\rho_0}$ represents the initial state and $\kket{0,0}^{\otimes k} = \kket{0}^{\otimes 2k}$ comes from the definition of the random variable $\omega$.  Each gate $E_\lambda$ is drawn independently and uniformly with respect to the Haar measure from an isometry group $\mathcal{E}$, which  can be either unitary or orthogonal~\footnote{
For the symplectic unitary group, the action on multiple copies of the system state simplifies to that of the standard unitary group. This is because the projector $|\pmb{x}\rangle\langle \pmb{x}|^{\otimes k}$ resides in the symmetric subspace, see for instance Ref.~\cite{west2024randomensemblessymplecticunitary}.}. 
This computation reduces to that of the replica transfer matrix 
\begin{equation}
    \mathcal{T}_\lambda\equiv \Ex_{E_\lambda \sim \mathcal{E}}[(E_\lambda \otimes E_\lambda^*)^{\otimes k}].\label{eq:tmat}
\end{equation}
Let us denote $q=d^{|\lambda|}$ as the Hilbert space dimension where the action of $E_\lambda$ is non-trivial, and define $\mathrm{Comm}_k(\mathcal{E})$ the $k$-commutant of $\mathcal{E}$, which consists of all operators $W$ 
such that $[W,E^{\otimes k}]=0$ for any $E\in \mathcal{E}$. 
By Schur-Weyl duality, the replica transfer matrix can be expressed as
\begin{align}
    \mathcal{T}_\lambda = \sum_{\sigma, \tau \in \mathrm{\mathrm{Comm}_k(\mathcal{E})} }\Wg_{\sigma,\tau}^\mathcal{E}(q) \kket{\sigma}\bbra{\tau}\;,
\end{align}
where $\Wg_{\sigma,\tau}^\mathcal{E}(q)$ represents the Weingarten matrix, which is the pseudo-inverse of the Gram matrix $\G_{\sigma,\tau}^\mathcal{E}=\bbrakket{\sigma}{\tau}$. 
For the unitary group, the $k$-commutant is given by $\mathrm{Comm}_k(\mathcal{U}(q))=\{ |\pi\rrangle \;|\; \pi\in S_k\}$, which corresponds to the algebra representing the permutation group $S_k$ over the $k$-replica space~\footnote{Sometimes we use a slight abuse of notation and describe as $\sigma$ the commutant elements $|\sigma\rrangle$.}. 
On the other hand, for the orthogonal group, the $k$-commutant takes the form $\mathrm{Comm}_k(\mathcal{O}(q))=\{ |\pi\rrangle \;|\; \pi\in \mathfrak{B}_k\}$, where $\mathfrak{B}_k$ denotes the Brauer algebra associated with the set of pairings $H_{2k}\subset S_{2k}$ of $2k$ elements, see Ref.~\cite{Collins_2006,turkeshi2023paulispectrummagictypical,west2024realclassicalshadows,khanna2025randomquantumcircuitstimereversal} for a comprehensive discussion. 
The summation over either free index of the Gram matrix satisfies
\begin{equation}
\label{eq:sumgram}
    \sum_{\sigma \in \mathrm{Comm}_k(\mathcal{E})} \G_{\sigma,\tau}^\mathcal{E}(q) = 
    \prod_{m = 0}^{k-1}(q+f_\mathcal{E}(m)),
\end{equation}
where $f_\mathcal{E}(m)$ is a function of $m$, that depends on the chosen ensemble. Specifically, for the unitary group $f_\mathcal{E}(m) = m$, whereas for the orthogonal $f_\mathcal{E}(m) = 2m$. 
Setting $q=1$, corresponding to a system with no qudit, recast the number permutations of $k$ elements $|S_k|=k!$ and of pairings of $2k$ elements $|H_{2k}|=(2k-1)!!$. 
Similarly, for the Weingarten matrix, a summation over either free index satisfies
\begin{align}
\label{eq:sumweingarten}
    \sum_{\sigma \in \mathrm{Comm}_k(\mathcal{E})} \Wg_{\sigma,\tau}^\mathcal{E}(q) = 
     \prod_{m = 0}^{k-1}(q+f_\mathcal{E}(m))^{-1}.
\end{align}
These summations play a crucial role in simplifying the computations for random matrix product states and in formulating the replica tensor network numerical methods, which we  revisit in the following subsection.

\subsection{Random matrix product state (RMPS)}\label{sec:rmps}
Matrix product states (MPS) are a fundamental class of quantum states $\ket{\psi}$ represented by the wave function
\begin{equation}
    \ket{\psi} = \sum_{\substack{x_1,\dots, x_N \\ \alpha, \beta, ..., \gamma}} A^{(1)}_{\alpha}(x_1)A^{(2)}_{\alpha \beta}(x_2) \dots A^{(N)}_{\gamma}(x_N) \ket{x_1 x_2 \dots x_N} \, ,
\end{equation}
where $x_i \in \{0, 1, \dots, d-1\}$ are indices labeling the Hilbert space basis of dimension $d$ of qudit $i$, while $\alpha, \beta ... \gamma \in \{1, 2, \dots, \chi\}$ are auxiliary indices spanning a space of dimension $\chi$, the so-called \textit{bond dimension}~\cite{schollwock2011the}. The tensors $A^{(i)}_{\alpha \beta}(x_i)$ can be seen as $\chi \times \chi$ matrices dependent on the local qubit variable $x_i$. The state can be pictorially represented in the bulk as
\begin{equation}\label{eq:mps_bulk}
\begin{tikzpicture}[baseline=(current  bounding  box.center),scale=0.8]
\definecolor{mycolor}{rgb}{0.85, 0.83, 0.89}
    \foreach \x in {1,...,5}{
        \draw[thick, black] (1.1*\x,0.8) -- (1.1*\x,1.6);
    }
    \draw[line width=0.8mm, gray!60!black, dotted] (-0.3,0.8) -- (0.2,0.8);
    \draw[line width=0.8mm, gray!60!black] (0.2,0.8) -- (6.4,0.8);
    \draw[line width=0.8mm, gray!60!black, dotted] (6.4,0.8) -- (6.9,0.8);
    \foreach \x in {1,...,5}{
        \draw[thick, fill=mycolor, rounded corners=2pt] (1.1*\x-0.4,0.4) rectangle (1.1*\x+0.4,1.2);
        \pgfmathsetmacro{\y}{int(\x - 3)}
        \node[scale=0.4] at (1.1*\x,0.8) {\Large $A^{(i \ifnum\y<0 
            \y 
        \else
            \ifnum\y=0
                \,
            \else
                + \y 
            \fi
        \fi)}$};
    }
\end{tikzpicture}\;, 
\end{equation}
where links denote the physical Hilbert space and thick lines indicate contractions over the bond dimension $\chi$. 
Random Matrix Product States (RMPS) are defined by assigning an appropriate probability measure to the tensors. One common prescription is to take the $A^{(i)}$ to be equal to a Haar-random matrix $E^{(i)}\in\mathcal{E}(d\chi)$ applied to the local basis state $\ket{0}$~\cite{garnerone2010typicality,
garnerone2010statistical,PRXQuantum.2.040308,PRXQuantum.4.030330,cheng2024pseudoentanglementtensornetworks}. Here, $\mathcal{E}$ represents either the unitary group ($\mathcal{U}$) or orthogonal group ($\mathcal{O}$). Graphically, in the bulk, we have therefore 
\begin{equation}\label{eq:rmps_bulk}
\begin{tikzpicture}[baseline=(current  bounding  box.center),scale=0.8]
\definecolor{mycolor}{rgb}{0.85, 0.83, 0.89}
    \foreach \x in {1,...,5}{
        \draw[thick, black] (1.1*\x,0) -- (1.1*\x,1.6);
        \draw[thick, fill=white] (1.1*\x,0) circle (0.2);
        \node[scale=0.9] at (1.1*\x,-0.45) {$|0\rangle$};
    }
    \draw[line width=0.8mm, gray!60!black, dotted] (-0.3,0.8) -- (0.2,0.8);
    \draw[line width=0.8mm, gray!60!black] (0.2,0.8) -- (6.4,0.8);
    \draw[line width=0.8mm, gray!60!black, dotted] (6.4,0.8) -- (6.9,0.8);
    \foreach \x in {1,...,5}{
        \draw[thick, fill=mycolor, rounded corners=2pt] (1.1*\x-0.4,0.4) rectangle (1.1*\x+0.4,1.2);
        \pgfmathsetmacro{\y}{int(\x - 3)}
        \node[scale=0.4] at (1.1*\x,0.8) {\Large $E^{(i \ifnum\y<0 
            \y 
        \else
            \ifnum\y=0
                \,
            \else
                + \y 
            \fi
        \fi)}$};
    }
\end{tikzpicture}\;.
\end{equation}
This construction allows to represent the state $\ket{\psi}$ via a suitable quantum circuit. In fact, we can reshape Eq.~\eqref{eq:rmps_bulk} into a staircase, where gates are sequentially ordered and act over $r+1$ sites, with $r\equiv \log_d(\chi)$~\cite{lami2024quantum,Lami_2023_1}
\begin{equation}\label{eq:rmpsOBC}
   \begin{tikzpicture}[baseline=(current  bounding  box.center),scale=0.6]
   \definecolor{mycolor}{rgb}{0.85, 0.83, 0.89}
    \foreach \x in {1,...,6}{
        \draw[thick, black] (\x,0) -- (\x,6);
        \draw[thick, fill=white] (\x,0) circle (0.2);
        \draw[line width=0.8mm, gray!60!black] (\x,\x+0.2-1) -- (\x,\x);
        \node[scale=0.75] at (\x,-0.6) {\large $|0\rangle$};
    }
    \foreach \x in {1,...,5}{
        \draw[thick, fill=mycolor, rounded corners=2pt] (\x-0.2,\x-0.4) rectangle (\x+1+0.2,\x+0.4);
        \node[scale=0.4] at (\x+0.5,\x) {\Large 
    $E^{\left( 
        \ifnum\x<3 
            \x 
        \else
            \ifnum\x=3\relax
                \, \dots
            \else
                \ifnum\x=4\relax
                    N - r - 1
                \else
                    N - r 
                \fi
            \fi
        \fi
    \right)}$
};

    }
\end{tikzpicture}\;.
\end{equation}

Finally, in the following, we will consider the ensemble of Gaussian random matrix product states. 
This ensemble is defined relaxing the unitarity condition and assuming that all MPS tensors $A^{(i)}_{\alpha \beta}(x_i)$ follow a Ginibre distribution, i.e., they have i.i.d. complex Gaussian entries with mean $0$ and a fixed variance $\nu^2$~\cite{lancien2021correlation}.
Although this approach does not produce normalized states $\ket{\psi}$, we will show that the ensemble of Gaussian RMPS reproduces the phenomenology of Haar unitary RMPS, given a sufficiently large  $\chi $ and an appropriately chosen $ \nu $. 
The key advantage of using Ginibre gates is that they enable an analytical treatment of more complex architectures, including brickwork circuits.

\subsection{Brickwork quantum circuits and replica tensor networks}
\label{subsec:bwcircuit}
Complementarily, we study the case of brickwork circuits (BW) where the gate application pattern alternates between even and odd time steps, respectively $\Lambda_s = \{(1,2),(3,4), ...,(N-1, N)\}$ for even depth and $\Lambda_s = \{(2,3),(4,5), ...,(N-2, N-1)\}$ for odd depth, cf. Eq.~\eqref{eq:circuit}. 
Graphically, this architecture is represented by 
\begin{equation}\label{eq:compact_vertical_brickwork}
   \begin{tikzpicture}[scale=0.8]
    \definecolor{mycolor}{rgb}{0.85, 0.83, 0.89}
    \def\n{8}
    \foreach \x in {1,...,\n}{
        \draw[thick, fill=white] (\x,-0.2) circle (0.2);
        \draw[thick] (\x,0) -- (\x,4.5);
        \node[below] at (\x,-0.4) {\large $|0\rangle$};
    }
    \foreach \x in {1,3,5,7}{
        \draw[thick, fill=mycolor, rounded corners=2pt] (\x-0.2,0.5) rectangle (\x+1.2,1.1);;
    }
    \foreach \x in {2,4,6}{
        \draw[thick, fill=mycolor, rounded corners=2pt] (\x-0.2,1.5) rectangle (\x+1.2,2.1);;
    }
    \foreach \x in {1,3,5,7}{
        \draw[thick, fill=mycolor, rounded corners=2pt] (\x-0.2,2.5) rectangle (\x+1.2,3.1);;
    }
    \foreach \x in {2,4,6}{
        \draw[thick, fill=mycolor, rounded corners=2pt] (\x-0.2,3.5) rectangle (\x+1.2,4.1);;
    }
   \end{tikzpicture}\nonumber
\end{equation}
where each two qubits gate is independently and identically drawn randomly from the ensemble $\mathcal{E}=\mathcal{U},\; \mathcal{O}$. 

Upon contracting with the state $|\pmb{0}\rangle=|0\rangle^{\otimes N}$ and taking the average, Eq.~\eqref{eq:tmat} specializes to the two qudit transfer matrix $\mathcal{T}^{(k)}_{i,i+1} \equiv \mathbb{E}_\mathrm{Haar}[(E_{i,i+1}\otimes E^\ast_{i,i+1})^{\otimes k}]$. 
Using the Weingarten calculus, we obtain
\begin{equation}
   \mathcal{T}^{(k)}_{i,i+1} = \sum_{\tau,\sigma\in \mathrm{Comm}_k(\mathcal{E})} \mathrm{Wg}^\mathcal{E}_{\tau,\sigma}(d^2) |\tau\rrangle_i |\tau\rrangle_{i+1}\llangle \sigma |_i \llangle \sigma|_{i+1}\;.
\end{equation}
Since the states $\kket{\tau}$ are \textit{not orthonormal} but, as anticipated,  $\bbrakket{\sigma}{\tau}=\G^\mathcal{E}_{\sigma,\tau}$, we conveniently reabsorb the overlaps by defining the tensors 
\begin{equation}
\begin{split}
	\mathcal{T}^{(k)}_{i,i+1} &\equiv 
	\begin{tikzpicture}[baseline=(current  bounding  box.center), scale=1]
\draw[thick] (-1.75,-0.2) -- (-1.75,0.4);
\draw[thick] (-1.25,-0.2)-- (-1.25,0.4);
\draw[thick, fill=myblue, rounded corners=2pt] (-1.9,0.3) rectangle (-1.1,-0.1);
\end{tikzpicture}
\equiv \sum_{\pi_1,\pi_2,\pi,\tau\in \mathrm{Comm}_k(\mathcal{E})} \mathrm{Wg}^\mathcal{E}_{\tau,\pi}(d^2) \times \\ & G^\mathcal{E}_{\pi,\pi_1}(d)G^\mathcal{E}_{\pi,\pi_2}(d) |\tau\rrangle_i |\tau\rrangle_{i+1}   \llangle \hat{\pi}_1|_i \llangle \hat{\pi}_2|_{i+1}\;,
\end{split}
\label{eq:wten}
\end{equation}
where we defined the dual states $|\hat{\sigma}\rrangle$ such that $\llangle \hat{\sigma}|\tau\rrangle = \delta_{\sigma,\tau}$. 
The first contraction of the replicated circuit can be simplified from the property $(\bbra{0,0}^{\otimes k}) \cdot \kket{\sigma} = 1$, which holds for any $\sigma\in \mathrm{Comm}_k(\mathcal{E})$ in both unitary and orthogonal ensembles. Using Eq.~\eqref{eq:sumweingarten}, we find that the first layer of replicated gates contracted with the replicated initial state $\kket{\rho_0}^{\otimes k}$ gives a product of
\begin{equation}
\begin{split}
	\begin{tikzpicture}[baseline=(current  bounding  box.center), scale=1]
\draw[thick] (-1.75,0) -- (-1.75,0.4);
\draw[thick] (-1.25,0)-- (-1.25,0.4);
\draw[thick, fill=myyellow, rounded corners=2pt] (-1.9,0.3) rectangle (-1.1,-0.1) (-1.5,0.08) node{$+$};
\end{tikzpicture}
&
=\sum_{\pi \in \mathrm{Comm}_k(\mathcal{E})} \frac{1}{\prod_{m=0}^{k-1}(d^2+f_\mathcal{E}(m))} |\pi\rrangle_{i}|\pi\rrangle_{i+1}\;.
\end{split}
\label{eq:initial}
\end{equation}
On the other hand, employing the definition of dual states we find that the contraction of the final layer of the replicated circuit with the final state $\llangle{0,0}|^{\otimes k}$ yields a product of 
\begin{equation}
\begin{split}
	\begin{tikzpicture}[baseline=(current  bounding  box.center), scale=1]
\draw[thick] (-1.75,-0.25) -- (-1.75,0.0);
\draw[thick] (-1.25,-0.25)-- (-1.25,0.0);
\draw[thick, fill=myorange, rounded corners=2pt] (-1.9,0.3) rectangle (-1.1,-0.1) (-1.5,0.08) node{$\hat{+}$};
\end{tikzpicture} \equiv \llangle \hat{+}|_i\llangle \hat{+}|_{i+1} \mathcal{T}^{(k)}_{i,i+1}\;,
\end{split}
\label{eq:lastLayer}
\end{equation}
with $\bbra{\hat{+}}=\sum_{\pi\in \mathrm{Comm}_k(\mathcal{E})} \bbra{\hat{\pi}}$. 
Summarizing, the computation of the average inverse participation ratios in brickwork circuits reduces to the \textit{replica tensor network} (RTN) contraction 
\begin{equation}
    I_k^{\mathrm{BW},\mathcal{E}} =  
    \begin{tikzpicture}[baseline=(current  bounding  box.center), scale=0.45]
    \draw [decorate,decoration={brace},thick] (-14,-6) -- node[left]{$t$}(-14,+.3);
  \foreach \i in {1,...,3} {
   \draw[thick, fill=myorange, rounded corners=1pt]  (-1.4-4*\i,+0.3) rectangle (1.4-4*\i,-0.7) (1.4-1.33-4*\i,-0.2) node{$\hat{+}$};
   }
  \foreach \kk[evaluate=\kk as \k using 0.25*\kk] in {0} {
  \pgfmathsetmacro{\col}{ifthenelse(int(mod(\kk,2))==0,"myblue","myblue")}
  \foreach \i in {1,...,6}{
    \draw[ thick] (-1.-2*\i+\k,-0.7+\k) -- (-1.-2*\i+\k,-5+\k);
  }
  \foreach \jj[evaluate=\jj as \j using -(ceil(\jj/2)-\jj/2), evaluate=\j as \fin using 3+\j] in {1,...,3} {
    \foreach \i in {1,...,\fin}{        
      \draw[thick, fill=\col, rounded corners=1pt] (-1.4-4*\i+4*\j,+0.5-1.3*\jj-0.25) rectangle (1.4-4*\i+4*\j,-0.5+-1.3*\jj-0.25);
    }
  }
  \foreach \jj[evaluate=\jj as \j using -(ceil(\jj/2)-\jj/2), evaluate=\j as \fin using 3+\j] in {4} {
    \foreach \i in {1,...,\fin}{        
      \draw[thick, fill=myyellow, rounded corners=1pt] (-1.4-4*\i+4*\j+\k,+0.5-2*\jj+2.5) rectangle (1.4-4*\i+4*\j+\k,-0.5-2*\jj+2.5) (1.4-1.33-4*\i+4*\j+\k,-0.5-2*\jj+0.5+2.5)node{$+$};
    }
  }
} \end{tikzpicture}\;.
\label{eq:Ydtensor}
\end{equation}

\section{Anticoncentration of Haar and random matrix product ensembles}

We are now in a position to discuss our analytical and numerical results. 
After briefly revisiting the distribution of overlaps for unitary and orthogonal Haar ensembles, we proceed to compute the anticoncentration properties of random matrix product states. This analysis enables us to identify the universal structure of the leading, subleading, and sub-subleading coefficients. 
We conjecture that this form is universal across all chaotic many-body systems, subject to the symmetries of the problem, such as time-reversal invariance~\cite{turkeshi2023paulispectrummagictypical,khanna2025randomquantumcircuitstimereversal,west2024randomensemblessymplecticunitary}.

\subsection{Anticoncentration of Haar ensembles}
\label{subsec:AnticoncentrationHaar}
We begin by briefly recalling the anticoncentration properties of random Haar states. These states are generated by applying a global operation, $E=E_{\{1,\dots,N\}}\in \mathcal{E}(d^N)$, to the many-body reference state $|\pmb{0}\rangle = |0\rangle^{\otimes N}$, where the ensemble $\mathcal{E}$ can be either $\mathcal{U}$ or $\mathcal{O}$. After recasting the inverse participation ratios as in Eq.~\eqref{eq:IPR_circuit}, with $E_t = E$ and $\mathcal{D} = \mathcal{E}$, we can employ the identity $(\bbra{0,0}^{\otimes k}) \cdot \kket{\sigma} = 1$ for any $\sigma \in \mathrm{Comm}_k(\mathcal{E})$, along with the Weingarten expression in Eq.~\eqref{eq:sumweingarten}, as derived in~\cite{Lami2025}, to obtain 
\begin{equation}
I_k^{\mathrm{Haar},\mathcal{E}}\equiv \mathbb{E}_{U\sim\mathcal{E}}[I_k(U|\pmb{0}\rangle)] = D\frac{{\prod_{m=0}^{k-1} (1+f_\mathcal{E}(m))}}{\prod_{m=0}^{k-1} (D+f_\mathcal{E}(m))}\;,
\end{equation}
where $f_\mathcal{E}(m)$ is determined by the ensemble, see Sec.~\ref{subsec:wmethodsa}, and the numerator corresponds to $|\mathrm{Comm}_k(\mathcal{E})|$. Thus, the explicit form is
$I_k^{\mathrm{Haar, \mathcal{U}}}= D \frac{k!}{\prod_{m=0}^{k-1} (D+m)}$ for the Unitary ensemble, and $I_k^{\mathrm{Haar, \mathcal{O}}}= D \frac{(2k-1)!!}{\prod_{m=0}^{k-1} (D+2m)}$.

From the expression of $I_k^{\mathrm{Haar},\mathcal{E}}$, we can compute the generating function for the stochastic variable $\omega$, cf. Sec.~\ref{sec:methods}, which is given by 
\begin{equation}
    \tilde{\mathcal{P}}_\mathcal{E}(x)\equiv \sum_{k=0}^\infty D^{k-1} I_k^{\mathrm{Haar},\mathcal{E}}\frac{(-x)^k}{k!}\;,
\end{equation}
which can be resummed for both unitary and orthogonal ensembles as $\tilde{\mathcal{P}}_\mathcal{U}(x)= (D-1)\int_{0}^{1} t^{D-2}e^{-Dx(1-t)}dt$ and $\tilde{\mathcal{P}}_\mathcal{O}(x)= \frac{\Gamma(D/2)}{\sqrt{\pi}\Gamma((D-1)/2)}\int_0^1 \frac{(1-t)^{\frac{D-3}{2}}}{\sqrt{t}}e^{-Dxt}dt$, where $\Gamma(x)$ is the gamma function. We can compute the inverse Laplace transform $\mathcal{L}^{-1}\{\tilde{\mathcal{P}}_\mathcal{E}\}$ to get the distribution, using $\mathcal{L}^{-1}\{x\mapsto e^{-ax}\}(\omega)=\delta(\omega-a)$. We obtain $\mathcal{P}_\mathcal{U}(\omega)= \frac{D-1}{D}(1-\frac{\omega}{D})^{D-2}$ and $\mathcal{P}_\mathcal{O}(\omega)= \frac{\Gamma(D/2)}{\sqrt{D}\Gamma((D-1)/2)}\frac{1}{\sqrt{\pi \omega}}(1-\frac{\omega}{D})^{(D-3)/2}$. 

In the limit $D\gg 1$, the Porter-Thomas distribution for the unitary ensemble reduces to the exponential distribution
\begin{equation}
\begin{split}
    I_k^{\mathrm{Haar},\mathcal{U}} &= \frac{k!}{D^{k-1}}\;,\quad \mathcal{P}_\mathcal{U}(\omega)= e^{- \omega}\;.
\end{split}
\end{equation}
On the other hand, for the orthogonal ensemble, it follows a chi-squared distribution~\cite{porter1956}
\begin{equation}
    I_k^{\mathrm{Haar},\mathcal{O}} = \frac{(2k-1)!!}{D^{k-1}}\;,\quad \mathcal{P}_\mathcal{O}(\omega)= \frac{1}{\sqrt{2\pi \omega}}e^{-\frac{\omega}{2}}\;.
\end{equation}
When the ensemble is clear from the context, we simplify the notation by writing $I_k^{\mathrm{Haar},\mathcal{E}}\mapsto I_k^{\mathrm{Haar}}$. 

\subsection{Anticoncentration in RMPS}
\label{subsec:rmps anticon}

We start by revisiting the results of Ref.~\cite{Lami2025}, which demonstrate that for $\chi\gg N$ the IPRs of RMPS converge to those of the Haar ensemble. A specific scaling limit has been also identified, appearing when the ratio $N/\chi$ is kept fix for $N \rightarrow \infty$. In this limit, we have been able to write the overlap probability distribution $\mathcal{P}(\omega)$, which depends on the value of the ratio. Here, we extend this calculation by introducing finite size $N$ corrections to the distribution.

By considering the case of unitary RMPS, the computation of IPRs involves a replica circuit, constructed from Eq.~\eqref{eq:rmpsOBC} with an additional contraction with all zeroes at the end, namely
\begin{equation}
    \begin{tikzpicture}[baseline=(current  bounding  box.center),scale=0.8]
       \definecolor{mycolor}{rgb}{0.85, 0.63, 0.39}
       \foreach \x in {1,...,6}{
           \draw[thick, black] (\x,0) -- (\x,6);
           \draw[thick, fill=white] (\x,0) circle (0.2);
    
           {\draw[line width=0.8mm, red!60!black] (\x,\x+0.2-1) -- (\x,\x);} 
    
           \draw[thick, fill=white] (\x,6) circle (0.2);
           \ifthenelse{\x=1}
            {\node[scale=0.6] at (\x,-0.6) {\large $|0,0\rrangle^{\otimes  k r}$};
           }
           {{\node[scale=0.6] at (\x,-0.6) {\large $|0,0\rrangle^{\otimes  k }$};
           }}
           \ifthenelse{\x=6}
           {\node[scale=0.6] at (\x,6.6) {\large $\bbra{0,0}^{\otimes k r}$};
           }
           {\node[scale=0.6] at (\x,6.6) {\large $\bbra{0,0}^{\otimes k }$};
           }}
       \foreach \x in {1,...,5}{
           \draw[thick, fill=mycolor, rounded corners=2pt] (\x-0.2,\x-0.4) rectangle (\x+1+0.2,\x+0.4);
           \node[scale=0.6] at (\x+0.5,\x) {\Large $\mathcal{T}^{(k)}$};
       }
    \end{tikzpicture}
    \label{eq:replicaRMPS}
\end{equation}
where the gates are
\begin{equation}
    \mathcal{T}^{(k)} = \sum_{\tau,\sigma\in \mathrm{Comm}_k(\mathcal{E})} \mathrm{Wg}^\mathcal{E}_{\tau,\sigma}(d\chi) |\tau\rrangle_i\llangle \sigma |_i \;.
\end{equation}
As before, certain contractions with zeroes are trivial, leading to a free sum over Weingarten, i.e. Eq.~\eqref{eq:sumweingarten}. 
Meanwhile, the contraction of each red leg, which lives in the auxiliary dimensions, yields $\bbra{\sigma}\tau\rrangle$, corresponding to  $G_{\sigma,\tau}(\chi)$, summed over one index as in Eq.~\eqref{eq:sumgram}.
Applying this process to every gate we arrive at the final result
\begin{align}
\begin{split}
    \label{eq:exact IPR}
I^{\mathrm{RMPS},\mathcal{E}}_{k}&= D\prod_{m=0}^{k-1}\left(\frac{1+f_\mathcal{E}(m)}{d\chi+f_\mathcal{E}(m)}\right)\left[\prod_{m=0}^{k-1}\left(\frac{\chi+f_\mathcal{E}(m)}{d\chi+f_\mathcal{E}(m)}\right)\right]^{N-r-1}.
    \end{split}
\end{align}
As anticipated, we now consider the scaling limit $N\to\infty$ while keeping 
\begin{equation}
   x_{\rm RMPS}=\frac{N}{\chi}\frac{d-1}{d} 
\end{equation}
constant. In this limit, we simplify Eq.~\eqref{eq:exact IPR} and identify the deviations from the Haar value up to order $O(1/N)$, as follows
\begin{equation}
\begin{split} 
\frac{I^{\mathrm{RMPS},\mathcal{U}}_{k}}{I^{\mathrm{Haar},\mathcal{U}}_{k}} &= \displaystyle e^{\frac{k(k-1)}{2}\alpha_{\rm RMPS}}e^{-k(k-1)(k-1/2)\beta^\mathcal{U}_{\rm RMPS}}+\ldots\;,\label{eq:ziocan}\\
    \frac{I^{\mathrm{RMPS},\mathcal{O}}_{k}}{I^{\mathrm{Haar},\mathcal{O}}_{k}} &= \displaystyle e^{{k(k-1)}\alpha_{\rm RMPS}}e^{- k(k-1)(k-1/2)\beta^\mathcal{O}_{\rm RMPS}}+\ldots\;.
\end{split}
\end{equation}
In the above expression, the scaling variables $\alpha_{\rm RMPS}$ and $\beta^\mathcal{E}_{\rm RMPS}$ are given by 
\begin{equation}
\label{eq:alphabeta}
    \begin{split}
       &  \alpha_{\rm RMPS} =  x_{\rm RMPS}\left(1-\frac{d}{N(d-1)} -\frac{\log_d[N (d-1)/xd]}{N}\right)\;,\\ & 
        \beta^\mathcal{U}_{\rm RMPS} =\frac{ x_{\rm RMPS}^2}{6N}\frac{d+1}{d-1} \, , \qquad  \, \, \beta^\mathcal{O}_{\rm RMPS}=2\frac{ x_{\rm RMPS}^2}{3N}\frac{d+1}{d-1} \, .
    \end{split}
\end{equation}
The $1/N$ terms are finite size (or depth) corrections to the scaling limit, and we shall see later that they can play a strong role in fixing the correct distribution of overlaps or for benchamarking the output of the quantum circuit. If we omit these corrections, by applying Eq.~\eqref{eq:ziocan} and following a similar approach to Refs.~\cite{Lami2025,christopoulos2024universaldistributionsoverlapsunitary}, we can now express the overlap $\omega$ as a product of two independent random variables $\omega =\omega_1 \omega_2$. Here, $\omega_1 \sim \mathcal{P}_\mathrm{PT}(\omega_1)$ is Porter-Thomas distributed, while $\omega_2 \sim \mathcal{P}_\mathrm{LN}(\omega_2)$ follows the Lognormal distribution
\begin{equation}
    \mathcal{P}_\mathrm{LN}(\omega) = \frac{1}{\omega \sigma \sqrt{2\pi}} \exp\left( -\frac{(\ln \omega - \mu)^2}{2\sigma^2} \right), \quad \omega > 0\;.
\end{equation}
Hence, the distribution of $\omega$ can be expressed as a suitable convolution of the two via convolution
\[
\mathcal{P}(\omega) = \int_{-\infty}^{\infty} \frac{1}{|\omega_2|} \, \mathcal{P}_\mathrm{PT}\left( \frac{\omega}{\omega_2} \right) \mathcal{P}_\mathrm{LN}(\omega_2) \, d\omega_2.
\]
Performing simple changes of variables, and setting  $\mu$ and $\sigma^2$ in the LogNormal distribution by matching of the moments $\exp[k(k-1)/2\alpha] = \exp[k\mu + k^2 \sigma^2/2]$, we obtain
\begin{equation}
\begin{split}
\mathcal{P}^{\mathcal{U}}_0(\omega;\alpha)&\equiv\int_{-\infty}^{+\infty}\frac{du}{\sqrt{2\pi}}e^{-\frac{u^2}{2}+\alpha}e^{-\omega e^{u\sqrt{\alpha}+\frac{3}{2}\alpha}}\;,\\
\mathcal{P}^{\mathcal{O}}_0(\omega;\alpha)&\equiv\int_{-\infty}^{+\infty}\frac{du}{\sqrt{2\pi^2 \omega}}e^{-u^2+\frac{3}{4}\alpha}e^{-\frac{\omega}{2}e^{2 u\sqrt{{\alpha}}+2\alpha}}\;.
\end{split}
\end{equation}
If we now want to take the subsubleading order terms into account, we can perturbatively expand the overall distribution to first order in $\beta_{\rm RMPS}$. This leads to the following result:
\begin{align}
    \label{eq:Haar conjecture distribution}
    \mathcal{P}^{\mathrm{RMPS}}(\omega;\mathcal{E}) =  \bigg( 1 &+ \beta^\mathcal{E}_{\rm RMPS} \big[ 3 + 12\omega \partial_\omega + \nonumber \\
    & + \frac{15}{2} \omega^2 \partial_\omega^2 + \omega^3 \partial_\omega^3 \big] \bigg) \mathcal{P}^{\mathcal{E}}_0(\omega;\alpha_{\rm RMPS}) \; ,
\end{align}
where $\partial_\omega^{n}\mathcal{P}$ denotes the $n$-th derivative of $\mathcal{P}$.

As elaborated in the following section, we conjecture that the functional form of the subsubleading term in Eq.~\eqref{eq:ziocan}, carrying the coefficients $\beta$, is universal. We expect that the overlap distribution for a wide range of models will conform to the structure outlined in Eq.~\eqref{eq:Haar conjecture distribution} with the microphysics fixed only by the (generic) parameters $\alpha$ and $\beta$. 
To support this conjecture, the next section presents a heuristic description based on domain walls in the statistical models derived from Haar averages~\cite{zhou2020,nahum2017quantum,zhou_emergent_2019,Fisher2023}. 
We then derive analytical results in the random phase model (RPM), which exhibits the same universal structure.
Later, we test our assumptions through extensive numerical simulations, robustly corroborating our physically motivated hypothesis. 

\section{Universal form of finite size/depth corrections}

In this section, we promote our conjecture about the universal functional form of the finite-size corrections to the IPR. We start by presenting an analytic argument based on the domain walls picture, which extends our results for the RMPS case. Afterward, we present exact results for the random phase model, which further confirm our hypothesis.

\subsection{The domain walls picture and the RMPS cases}
While the results derived in the previous section, such as Eq.~\eqref{eq:Haar conjecture distribution}, are exact, they do not provide an intuitive or general understanding of anticoncentration across arbitrary circuit architectures. 
For this scope, we introduce instead an \textit{effective description based on domain walls}, which arise by interpreting the contraction of the replica tensor network in Eq.~\eqref{eq:Ydtensor} as the partition function of a suitable statistical mechanics problem. In this picture, permutations play the role of Ising-like  variables~\cite{zhou2020,nahum2017quantum,zhou_emergent_2019,Fisher2023}.  
The resulting model is ferromagnetic, with a coupling strength that increases over time (or circuit depth), and features multiple degenerate ground states. 
By studying domain wall excitations above these ground states and performing a strong-coupling expansion at large times, we derive a general expression for the inverse participation ratio (IPR) that captures both the leading behavior and finite-size corrections.

Crucially, our analysis is expected to hold quite generally for chaotic local random circuits. The results apply in the limit of large system size and sufficiently long times, where universality emerges and our expansions are valid.

We consider a random quantum circuit $C$ of depth $t$. We consider the evaluation of the $k-$th moment of the overlap $\omega$ with the computational basis, i.e.\  $I_k \propto \Ex[\omega^k]$.
Let $T_i$ denote the collection of all gates acting within a slice of $C$ at a fixed spatial position (site) $i$. We assume that the matrices $T_i$ are independent across different sites, as naturally arises from a circuit $C$ composed of local, independently and i.i.d. gates. The size of $T_i$, denoted $M(t)$, grows exponentially with $t$. The $T_i$ acts as a transfer matrix in the \textit{spatial direction}, as discussed earlier in Sec.~\ref{subsec:rmps anticon} (cf. Eq.~\eqref{eq:replicaRMPS} and Ref.~\cite{bertini2020scrambling}). 
Therefore, we can write the overlap as $\omega=|l^\dagger T_1 T_2 \cdots T_N r|^2$, where $l$, $r$ are suitable boundary vectors. 
At large times and system sizes, where universality is expected to emerge, we coarse-grain the model by grouping $L$ consecutive sites together, thereby effectively partitioning the system into $\tilde{N} = N/L$ blocks. 
In general, we allow the coarse-graining length to vary with circuit depth, $L = L(t)$. This procedure allows us to express the overlap as
$\omega=|l^\dagger \Tilde{T}_0 \Tilde{T}_1 \cdots \, \Tilde{T}_{\tilde{N} -1} r|^2$, where each coarse-grained matrix is defined by $\Tilde{T}_a = T_{aL+1}\cdots \, T_{(a+1)L}$. 

We now introduce a crucial simplification by assuming that the coarse-grained matrices $\tilde{T}_i$ can be effectively treated as complex-valued Gaussian random matrices (i.e.\ Ginibre matrices). 
A similar assumption has been made in Ref.~\cite{christopoulos2024universaldistributionsoverlapsunitary}, motivated by the observation that the matrices $\tilde{T}_i$ are generally neither unitary nor Hermitian. Therefore, for sufficiently large $L$, it is natural to model them as drawn effectively from the simplest class of non-Hermitian random matrices—the Ginibre ensemble.
Moreover, we observe that for Gaussian RMPS, this property holds by construction (see Section~\ref{sec:rmps}). In this setting, the matrix dimension can be identified with the MPS bond dimension, $M = \chi$, and the coarse-graining length is simply $L = 1$. Note that in this case there is no notion of time $t$, although the bond dimension $\chi$ plays a role analogous to the exponential of $t$.

Under the assumption that the coarse-grained matrices $\tilde{T}_i$ are i.i.d.  complex Gaussian random variables with mean zero and variance $\nu^2$, the Weingarten matrix becomes homothetic with ratio $\nu^{2k}$, and the transfer matrix reduces to the Gram matrix, given by $G_{\sigma,\pi} = \llangle \sigma | \pi \rrangle$. As a result, the IPR is given by  
\begin{equation}\label{eq:ipr_ginibre}
I_k = D \, \nu^{2k\Tilde{N}} \, (l^\dagger)^{\otimes k} G^{\Tilde{N}-1} r^{\otimes k} \,.
\end{equation}
For simplicity we assume the boundary vectors to be $l=r=(1,1, \dots, 1)$~\footnote{For Gaussian RMPS, these boundary vectors are obtained by contracting the first and last tensors, $A^{(1)}$ and $A^{(N)}$, with the state $\ket{0}$ living in the bond dimension space. In fact, the contraction with a permutation $\sigma$ always gives $(\llangle 0,0 |^{\otimes  
 k} ) | \sigma \rrangle = 1$.}. 
By identifying permutations $\sigma \in S_k$ with spins having $k!$ levels, Eq.~\eqref{eq:ipr_ginibre} can be interpreted as the partition function for an Ising-like spin chain of length $\tilde{N}$~\cite{zhou2020}. The interaction between neighboring sites is described by the Gram matrix $G$, which exhibits a \textit{ferromagnetic nature}. This is because permutations that are ``close'' to each other have larger overlaps. Specifically, we can express the matrix elements as $G_{\sigma \pi} = M^{k-d(\sigma,\pi)}$, where $d(\sigma,\pi)$ represents the transposition distance between the two permutations $\sigma$ and $\pi$. We can now proceed to expand the IPRs in Eq.~\eqref{eq:ipr_ginibre} for large $M$ (i.e. large $t$). First, we write 
\begin{equation}\label{eq:transfmatginibre}
  G_{\sigma \pi} = M^k \left( \delta_{\sigma \pi} + \frac{1}{M} A_{\sigma \pi}^{(1)} + \frac{1}{M^2}A_{\sigma \pi}^{(2)} + o(M^{-2}) \right) \, ,  
\end{equation}
where $A^{(n)}$ is a matrix which connects permutations at distance $n$, i.e.\ $A^{(n)}_{\sigma \pi} = \delta_{n, \, d(\sigma,\pi)}$.
Retaining only the leading (diagonal) contribution $\delta_{\sigma \pi}$ in each of the matrices $G$ in Eq.~\eqref{eq:ipr_ginibre}, leads to a free sum over the $k!$ permutations. In the language of the spin model, these can be seen as $k!$ degenerate ferromagnetic ground states labeled by $\pi$, which we represent as follows:
\begin{equation}
\begin{tikzpicture}[baseline=(current  bounding  box.center),scale=0.8]
\definecolor{mycolor1}{rgb}{0.0, 0.83, 0.39}
\pgfmathsetmacro{\ll}{2.5}
\draw[line width=0.8mm, mycolor1, ultra thick] (-\ll,0.) -- (+\ll,0.);
\node[scale=1.] at (\ll*5/6,-0.35) {$\pi$};
\end{tikzpicture}\; .
\end{equation}
This leading contribution gives: $I_k=D k!\nu^{2k\Tilde{N}}M^{k(\Tilde{N}-1)}$. 
The normalization of the state implies $I_1=1$, 
which fix the Gaussian variance to $\nu^2=d^{-\frac{N}{\Tilde{N}}}M^{-\frac{\Tilde{N}-1}{\Tilde{N}}}$. With this choice, we recover the Haar value of the IPRs: $I_k=I_k^{\mathrm{Haar},\mathcal{U}}=D^{1-k} k!$. Now, let us identify the first subleading contribution by replacing one of the matrices $\delta_{\sigma \pi}$ with $\frac{1}{M} A_{\sigma \pi}^{(1)}$. 
The matrix $A_{\sigma \pi}^{(1)}$ enables a permutation $\pi$ to transition to one of its nearest neighbors $\sigma$. In the language of spin systems, the insertion of $A^{(1)}$ creates therefore a \textit{domain walls} between two ferromagnetic states. This situation can be represented as follows
\begin{equation}
\begin{tikzpicture}[baseline=(current  bounding  box.center),scale=0.8]
\definecolor{mycolor1}{rgb}{0.0, 0.83, 0.39}
\definecolor{mycolor2}{rgb}{0.83, 0.0, 0.39}
\pgfmathsetmacro{\ll}{2.5}
\draw[line width=0.8mm, mycolor2, ultra thick] (-\ll,0.) -- (0,0.);
\draw[line width=0.4mm, black, dotted] (0,-0.3) -- (0,0.3);
\draw[line width=0.8mm, mycolor1, ultra thick] (0,0.) -- (+\ll,0.);
\node[scale=1.] at (\ll*5/6,-0.35) {$\pi$};
\node[scale=1.] at (-\ll*5/6,-0.35) {$\sigma$};
\end{tikzpicture}\;, 
\end{equation}
where the dotted line $\raisebox{0.4ex}{\begin{tikzpicture}[baseline=(current  bounding  box.center),scale=0.8]
\draw[line width=0.4mm, black, dotted] (-0.3,0.0) -- (0.3,0.0);
\end{tikzpicture}}$ is the domain wall.
Since there are ${k(k-1)}/{2}$ permutations at a distance $1$ from $\pi$, the correction to the Haar IPRs due to the creation of a single domain wall is given by 
\begin{equation}
I_k \simeq I_k^\mathrm{Haar} \left( 1 + \frac{\Tilde{N}-1}{M} \frac{k(k-1)}{2} \right) \, .
\end{equation}
Next, we consider the correction from multiple domain walls. First, placing two instances of the matrix $A_{\sigma \pi}^{(1)}$ creates two domain walls, as represented here
\begin{equation}
\begin{tikzpicture}[baseline=(current  bounding  box.center),scale=0.8]
\definecolor{mycolor1}{rgb}{0.0, 0.83, 0.39}
\definecolor{mycolor2}{rgb}{0.83, 0.0, 0.39}
\definecolor{mycolor3}{rgb}{0.0, 0.39, 0.83}
\pgfmathsetmacro{\ll}{2.5}
\draw[line width=0.8mm, mycolor3, ultra thick] (-\ll,0) -- (-\ll/3,0.);
\draw[line width=0.4mm, black, dotted] (-\ll/3,-0.3) -- (-\ll/3,0.3);
\draw[line width=0.8mm, mycolor2, ultra thick] (-\ll/3,0.) -- (+\ll/3,0.);
\draw[line width=0.4mm, black, dotted] (+\ll/3,-0.3) -- (+\ll/3,0.3);
\draw[line width=0.8mm, mycolor1, ultra thick] (+\ll/3,0.) -- (+\ll,0.);
\node[scale=1.] at (\ll*5/6,-0.35) {$\pi$};
\node[scale=1.] at (0,-0.35) {$\rho$};
\node[scale=1.] at (-\ll*5/6,-0.35) {$\sigma$};
\end{tikzpicture}\;.
\end{equation}
This contributes a factor 
\begin{equation}\label{eq:two_dws}
\frac{1}{M^2} \frac{(\Tilde{N}-1)(\Tilde{N}-2)}{2} \left( \frac{k(k-1)}{2} \right)^2 \, ,
\end{equation}
since there are $(\Tilde{N}-1)(\Tilde{N}-2)/{2}$ ways to place two domain walls in different positions. Second, placing a single $A_{\sigma \pi}^{(2)}$ matrix at one of the $\Tilde{N}-1$ sites introduce an additional correction of the same order $M^{-2}$. This is represented as follows: 
\begin{equation}\label{eq:doublejumps}
\begin{tikzpicture}[baseline=(current  bounding  box.center),scale=0.8]
\definecolor{mycolor1}{rgb}{0.0, 0.83, 0.39}
\definecolor{mycolor2}{rgb}{0.83, 0.0, 0.39}
\definecolor{mycolor3}{rgb}{0.0, 0.39, 0.83}
\pgfmathsetmacro{\ll}{2.5}
\draw[line width=0.8mm, mycolor3, ultra thick] (-\ll,0) -- (+\ll/3,0.);
\draw[line width=0.8mm, black, dotted] (+\ll/3,-0.35) -- (+\ll/3,0.35);
\draw[line width=0.8mm, mycolor1, ultra thick] (+\ll/3,0.) -- (+\ll,0.);
\node[scale=1.] at (\ll*5/6,-0.35) {$\pi$};
\node[scale=1.] at (-\ll*5/6,-0.35) {$\sigma$};
\end{tikzpicture}\; ,
\end{equation}
where $\raisebox{0.4ex}{\begin{tikzpicture}[baseline=(current  bounding  box.center),scale=0.8]
\draw[line width=0.8mm, black, dotted] (-0.3,0.0) -- (0.3,0.0);
\end{tikzpicture}}$ is a sort of ``double jump'' domain wall. The combinatorial contribution corresponds to the number of permutations $\sigma$ at a fixed distance of 2 from a given permutation $\pi$, which is $\frac{3 k - 1}{4} \binom{k}{3}$~\cite{brualdi2012}. 
Thus, the second-order correction reads 
\begin{equation}\label{eq:single_double_dw}
\frac{1}{M^2} (\Tilde{N}-1) \frac{3 k - 1}{4} \binom{k}{3} \, .
\end{equation}
Combining all these contributions, we obtain 
\begin{align}
\label{eq:ziocan1}
I_k &\simeq I_k^{\mathrm{Haar}} \bigg( 1 + \frac{\Tilde{N}-1}{M} \frac{k(k-1)}{2} + \frac{1}{2} \left(\frac{\Tilde{N}-1}{M} \frac{k(k-1)}{2} \right)^2 \nonumber\\ &- \frac{\Tilde{N}-1}{M^2} \frac{k (k-1) (k-\frac{1}{2}) }{6} \bigg) \, .
\end{align}

\begin{figure*}[t!]
    \centering
    \includegraphics[width=\linewidth]{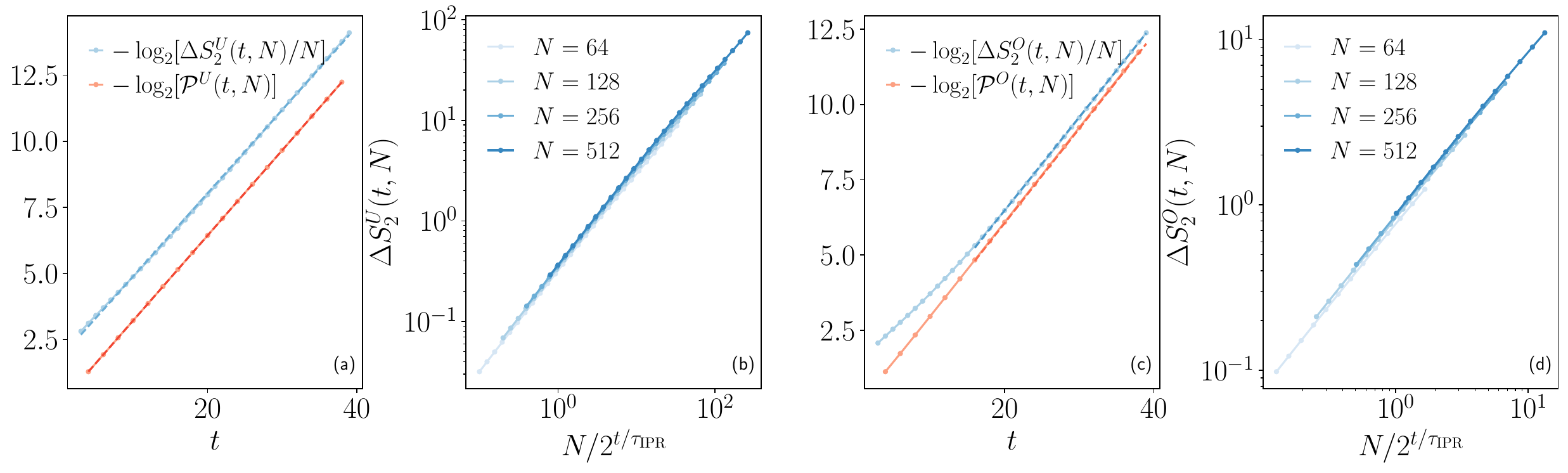}
    \caption{\footnotesize 
    (a) Scaling of $-\log_2[\Delta S_2^U(t,N)/N]$ and $-\log_2[\mathcal{P}^U(t,N)]$ for a brickwork random unitary circuit with $N=128$ qubits evolved up to time $t=40$. 
    The fits of the two curves (dashed lines) present the same slope with $\tau_{\mathrm{IPR}} = 3.108 \pm 0.002 \approx \tau_{\mathrm{PUR}} = 3.1063 \pm 0.0001$.
    (b) Using $ \tau_\mathrm{IPR} = 3.11$ we see a data collapse for $\Delta S_2^U(t,N)$ for     different circuit depth and system sizes $64 \leq N \leq 512$.
    (c) Scaling of $-\log_2[\Delta S_2^O(t,N)/N]$ and $-\log_2[\mathcal{P}^O(t,N)]$ for a brickwork random orthogonal circuit with $N=128$ qubits evolved up to time $t=40$. 
    We observe a change in the slope of the IPR around $t\simeq 16$ after which the scaling becomes approximately the same, $\tau_{\mathrm{IPR}} = 3.23 \pm 0.05$ and $\tau_{\mathrm{PUR}} = 3.19 \pm 0.01$. 
    (d) Data collapse for $\Delta S_2^O(t,N)$ imposing $\tau_{\mathrm{IPR}} = 3.2$ for different circuit depth and system sizes $64 \leq N \leq 512$. } 
    \label{fig:RTNplots}
    
\end{figure*}

We can now introduce the Thouless length $N_{\mathrm{Th}}(t)=M(t)L$, which reduces to $\chi$ for RMPS. 
Using the definitions $x=\frac{N}{N_{\mathrm{Th}}(t)}=\frac{\Tilde{N}}{M(t)}$ and $\alpha_{\rm RMPS} = x ( 1 - \frac{1}{\Tilde{N}}) = \frac{\Tilde{N}-1}{M}$, we can rewrite Eq.~\eqref{eq:ziocan1} as
\begin{align}
I_k & \simeq I_k^{\mathrm{Haar}} \bigg[ 1 + \alpha_{\rm RMPS} \frac{k(k-1)}{2} + \frac{1}{2} \left(\alpha_{\rm RMPS}\frac{k(k-1)}{2} \right)^2 \\ &-\frac{x^2}{6 \Tilde{N}} k (k-1) \left(k-\frac{1}{2}\right)  \bigg] \, .
\end{align}
If we continue the expansion up to terms of order $M^{-n}$, we will obtain contributions arising from placing $n$ domain walls at distinct positions, such for instance:
\begin{equation}\label{eq:n_dws}
\begin{tikzpicture}[baseline=(current  bounding  box.center),scale=0.8]
\definecolor{mycolor1}{rgb}{0.0, 0.83, 0.39}
\definecolor{mycolor2}{rgb}{0.83, 0.0, 0.39}
\definecolor{mycolor3}{rgb}{0.0, 0.39, 0.83}
\definecolor{mycolor4}{rgb}{1., 0.9, 0.}
\pgfmathsetmacro{\ll}{2.5}
\draw[line width=0.8mm, mycolor4, ultra thick] (-\ll,0) -- (-\ll/2,0.);
\draw[line width=0.4mm, black, dotted] (-\ll/2,-0.3) -- (-\ll/2,0.3);
\draw[line width=0.8mm, mycolor3, ultra thick] (-\ll/2,0.) -- (+0,0.);
\draw[line width=0.4mm, black, dotted] (+0,-0.3) -- (+0,0.3);
\draw[line width=0.8mm, mycolor2, ultra thick] (+0,0.) -- (+\ll/2,0.);
\draw[line width=0.4mm, black, dotted] (+\ll/2,-0.3) -- (+\ll/2,0.3);
\draw[line width=0.8mm, mycolor1, ultra thick] (+\ll/2,0.) -- (+\ll,0.);
\end{tikzpicture}\;.
\end{equation}
These contributions are analogous to Eq.~\eqref{eq:two_dws} (which corresponds to the case $n = 2$) and are of order $\tilde{N}^n / M^n$. This is because there are $\approx \tilde{N}^n$ possible ways to place the $n$ domain walls. However, additional contributions arise when two or more domain walls are placed at the same position, creating a domain wall between two permutations at a distance greater than $1$. Since there are fewer ways to place the domain walls when some of them coincide at the same point, these contributions have lower multiplicity (i.e. lower entropy), resulting in lower powers of $\tilde{N}$. For example, the term of Eq.~\eqref{eq:single_double_dw} is of order $\tilde{N}/M^2$, while Eq.~\eqref{eq:two_dws} is of order $\tilde{N}^2/M^2$.
By collecting all terms like Eq.~\eqref{eq:n_dws} at a generic order $n$, we can factor out a leading contribution which takes the form of an exponential. This finally leads to the results presented in
Eq.~\eqref{eq:ziocan}. 

\subsection{Another example: the random phase model}
In this section we corroborate our universality conjecture by studying an exactly solvable model: the Random Phase Model (RPM)~\cite{RPMPhysRevLett.121.060601}.
This quantum circuit model consists of $t$ layers alternating between single-site Haar unitaries $u^{(1)}_{i}$ and two-site random phase gates $[u^{(2)}_{i,i+1}]_{a_i,a_{i+1}}=\exp(i\varphi^{(j)}_{a_i,a_{i+1}})$ where the random phases $\varphi^{(j)}_{a_i,a_{i+1}}$ are drawn from a normal distribution $\varphi^{(j)}_{a_i,a_{i+1}}\sim \mathcal{N}(0,\epsilon)$ and $a_i\in\{1,...,d\}$. 
The parameter $\epsilon$ controls the strength of the gate  coupling. 
This model can be interpreted as a brickwork circuit as described in Sec.~\ref{subsec:bwcircuit}, with local gates 
 \begin{equation}\label{eq:RPM local gates}
   \begin{tikzpicture}[scale=0.8]
   \definecolor{mycolor}{rgb}{0.85, 0.83, 0.89}
    \def\n{5}
    
    \foreach \x in {1,2,4,5}{
        \draw[thick] (\x,-0.5) -- (\x,2.1);
    }

    \foreach \x in {1}{
        \draw[thick, fill=mycolor,,rounded corners=2pt] (\x-0.2,0.5) rectangle (\x+1.2,1.1);
    };
    
    \node[scale=0.9] at (3,0.7) {\huge =};
    \foreach \x in {4}{
        \draw[thick, fill=mycolor] (\x-0.2,0.5) rectangle (\x+1.2,1.1) (\x+2,0.8) node{$u^{(2)}_{i,i+1}$};
        \draw[thick, fill=mycolor,, rounded corners=2pt] (3.8,-0.2) rectangle (4.2,0.2) (\x-0.8,-0.2) node{$u^{(1)}_{i}$} ;
        \draw[thick, fill=mycolor,, rounded corners=2pt] (4.8,-0.2) rectangle (5.2,0.2)
        (\x+1.8,-0.2) node{$u^{(1)}_{i+1}$};
        \draw[thick, fill=mycolor,, rounded corners=2pt] (3.8,1.4) rectangle (4.2,1.8) (\x-0.8,1.9) node{$u^{(1)}_{i}$};
        \draw[thick, fill=mycolor,, rounded corners=2pt] (4.8,1.4) rectangle (5.2,1.8) (\x+1.8,1.9) node{$u^{(1)}_{i+1}$};
    }
    \node[scale=0.9] at (6.7,0.5) {,};

   \end{tikzpicture}\;.
   \nonumber
\end{equation}

We aim to compute the IPRs for this model. 
It turns out that taking the limit $d\xrightarrow[]{}+\infty$ while keeping the coupling $\epsilon$ fixed renders the contraction described in Eq.~\eqref{eq:Ydtensor} analytically tractable~\cite{christopoulos2024universaldistributionsoverlapsunitary}. 
In this limit, where $d\gg 1$, the single site Weingarten function becomes diagonal. 
As a result, the contribution from the unitaries to the transfer matrix simplifies as $\mathbb{E}_\mathrm{Haar}[(u^{(1)}_{i}\otimes u^{(1)\ast}_{i})^{\otimes k}] = \sum_{\tau,\sigma} \mathrm{Wg}^\mathcal{U}_{\tau,\sigma}(d) |\tau\rrangle_i \llangle \sigma |_i \sim d^{-k} \sum_{\tau\in S_k}|\tau\rrangle_i \llangle \tau |_i\;$. 
Additionally, both the Weingarten function for the random phase average and the Gram matrix $\G^\mathcal{U}_{\sigma,\tau}=\bbrakket{\sigma}{\tau}=d^{\#(\sigma^{-1}\tau)}$ become diagonal in the infinite dimension limit. 
Thus, the entire transfer matrix calculation boils down to evaluating the random phase average $\mathbb{E}[\llangle \sigma |\llangle \sigma'|(u^{(2)}_{i,i+1}\otimes u^{(2)\ast}_{i,i+1})^{\otimes k}|\sigma \rrangle |\sigma'\rrangle]$. 
This computation has been thoroughly analyzed in~\cite{christopoulos2024universaldistributionsoverlapsunitary} and, once again, simplifies significantly when considering only the leading term in $d$.
Taking all these contributions into account, we can now express the transfer matrix in permutation space as
 \begin{equation}
     \label{eq:TransferMatRPM}
     \left[m\right]_{\sigma\sigma'}=\begin{tikzpicture}[baseline=(current  bounding  box.center), scale=1]
\draw[thick] (-1.75,-0.4) -- (-1.75,0.6);
\draw[thick] (-1.25,-0.4)-- (-1.25,0.6);
\draw[thick, fill=myblue, rounded corners=2pt] (-1.9,0.3) rectangle (-1.1,-0.1);
\node at (-1.9,-0.5) {$\sigma$};
\node at (-1,-0.45) {$\sigma'$};
\node at (-1.9,0.7) {$\sigma$};
\node at (-1,0.75) {$\sigma'$};
\end{tikzpicture}=\exp\{-\epsilon (k-n_\mathrm{F}(\sigma\sigma'^{-1}))\},
 \end{equation}
where $n_\mathrm{F}(\sigma)$ denotes the number of fixed points of the permutation $\sigma$. 
Since the transfer matrix is diagonal in permutation space, we can perform the contraction of the $t$ layers of the circuit straightforwardly to get a transfer matrix in the spatial direction that is just $[\mathcal{T}_{\mathrm{RPM}}]_{\sigma\sigma'}=\left[m\right]_{\sigma\sigma'}^{\frac{t}{2}}$, where we assume $t$ is even. 
The IPRs can then be expressed as the product along the spatial direction  
\begin{align}
 \label{eq:IPR RPM}
     I_k^{\mathrm{RPM}}&=\frac{1}{D^{k-1}} \sum_{\sigma_1,...,\sigma_N\in S_k}\prod_{j=1}^{N-1} \left[\mathcal{T}_{\mathrm{RPM}}\right]_{\sigma_i\sigma_{i+1}} \nonumber\\&=\frac{1}{D^{k-1}} \llangle \hat{+}|\mathcal{T}_{\mathrm{RPM}}^{N-1}|+\rrangle\,.
 \end{align}
Although the transfer matrix differs from the one in Eq.~\eqref{eq:transfmatginibre}, the domain wall picture remains valid, with the only modification being the cost of each domain wall.
This cost depends on the Thouless length $N_{\mathrm{\mathrm{Th}}}(t)=e^{\epsilon t}$. 
In particular, the cost of a single domain wall is $1/N_{\mathrm{Th}}$ while the cost of a double domain wall at the same site is $1/N_{\mathrm{Th}}^{3/2}$. 
Consequently, the latter introduces $1/\sqrt{N}$ corrections to the IPRs rather than $1/N$.

By once again taking the scaling limit $N\rightarrow +\infty$ while keeping 
\begin{equation}
    x_{\rm RPM}=\frac{N}{N_{\mathrm{Th}}(t)}
\end{equation}
constant and accounting for all the domain walls configurations, we get finite-size corrections to the IPRs derived in~\cite{christopoulos2024universaldistributionsoverlapsunitary} 
\begin{equation} 
    \frac{I^{\mathrm{RPM}}_{k}}{I^{\mathrm{Haar}}_{k}} = \displaystyle e^{\frac{k(k-1)}{2} x_{\rm RPM}}\left(1-k(k-1)(k-2)\frac{ x_{\rm RPM}^{3/2}}{3\sqrt{N}}+O\left(\frac{1}{N}\right)\right)\;,\label{eq:RPM IPR}
\end{equation}
 together with the subleading term. The $(k-2)$ factor in the subleading term differs from the RMPS result \ref{eq:ziocan}, where instead is given by $(k-\frac{1}{2})$. This discrepancy arises in the domain wall picture from the different ways in which spin neighbors are defined compared to the Ginibre ensemble. In the Ginibre case, two permutations are considered $p$-neighbors if they differ by $p$ transpositions. In contrast, for the RPM, two permutations $\sigma$ and $\tau$ are deemed $p$-neighbors if the permutation $\sigma \tau^{-1}$ has $k-p$ fixed points. This distinction modifies the domain wall structure and accounts for the higher $1/\sqrt{N}$ corrections. 
 
 
\begin{figure*}
    \centering
    \includegraphics[width=\linewidth]{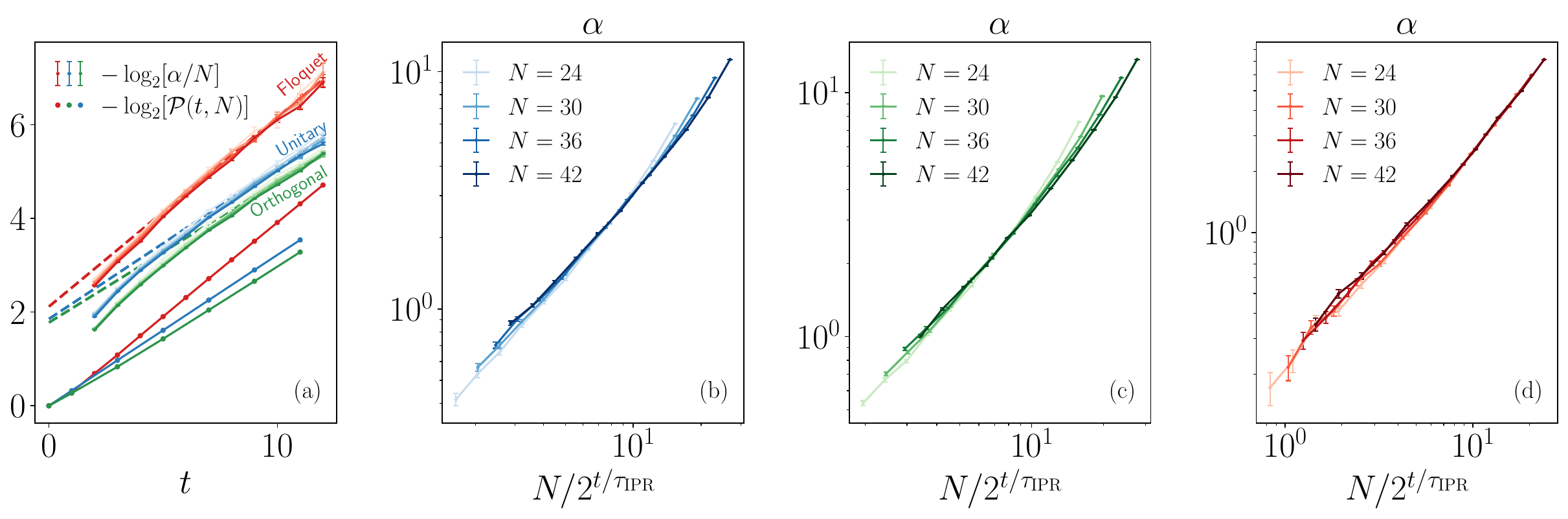}
    \caption{Panel \textbf{(a)} shows the scaling with time of $\alpha$ and the purity for the Floquet circuit and Unitary and Orthogonal brickwork circuits for $N=24, 30,36,42$ (darker shades correspond to higher $N$). We sampled the overlap distribution using tensor networks but without truncating the bond dimension. We find the best fit of $\alpha$ to this data (20 000 samples) through Maximum Likelihood Estimation.  In the Floquet and Unitary cases, both $\alpha$ and purity scale in a very similar fashion with $\tau_{\mathrm{IPR,F}}=2.47\pm 0.09$ and $\tau_{\mathrm{PUR,F}}=2.484 \pm 0.003$ while $\tau_{\mathrm{IPR,U}}=3.10\pm 0.11$ and $\tau_{\mathrm{PUR,U}}=3.108\pm 0.002$. The Orthogonal case reveals discrepancies between these values, with $\tau_{\mathrm{IPR,O}}=3.32\pm 0.05$ and $\tau_{\mathrm{PUR,O}}=3.21\pm 0.01$. These results are coherent with what was explained in Fig. \ref{fig:RTNplots}. Panels \textbf{(b)}, \textbf{(c)} and \textbf{(d)} show that $\alpha$ scales with $N/2^{t/\tau_{\mathrm{IPR}}}$ independently of $N$ for Unitary, Orthogonal and Floquet circuits respectively. The error bars indicate the standard deviation of the estimator.}
    \label{fig:plotsalpha}
\end{figure*}

\section{ Universal form of the distribution and numerical results}

{We are now in position to formulate a conjecture for the general form of the moments $I_k$ which can be applied to generic quantum circuits. }
 Remarkably, the dependence on the microscopic details of the system is fully captured by just two parameters, which we denote as $\alpha$ and $\beta$ and which we retain as fitting parameters. Later, we substantiate this conjecture through extensive numerical simulations, performed on both brickwork and Floquet quantum circuits. Finally, we show how our conjecture is useful also for extracting the global fidelity of a generic quantum circuit at small depth. 
\begin{conjecture}
\label{conjecture}
 States of $N$ qudits generated from generic quantum circuits evolved up to times $t \sim \log N$, exhibit the following general form for the inverse participation ratios
\begin{equation}\label{eq:conjec}
    I_{k} = I^{\mathrm{Haar},\mathcal{E}}_{k} \displaystyle e^{\frac{k(k-1)}{2}\alpha} e^{-k^2(k-1)\beta} + O(\beta^2) \, ,
\end{equation}
 where $\mathcal{E} \in \mathcal{O}, \mathcal{U}$ for orthogonal and unitary circuits, respectively, and $\alpha, \beta$ are system-dependent parameters.
These coefficients are expected to scale as 
\begin{equation}
\alpha= O(N/2^{t/\tau}), \quad     \beta = O( N/2^{\kappa t/\tau}),
\end{equation}
with two positive constants, $\tau$ and $\kappa>1$, determined by the microphysics of the circuit. 
\end{conjecture}

Our conjecture is motivated by the results for the RMPS and the random phase model presented in the previous section as well as the domain wall picture. 
{Indeed, according to the latter, the general form of the moments 
\begin{equation}
   I_k= I_k^{\rm Haar} e^{\frac{k(k-1)}{2}\alpha} e^{-k (k-1)(k-k_0)\beta}
\end{equation}
where $k_0$ is a model-dependent constant which depends on the combinatorials associated to the double jumps processes Eq.~\eqref{eq:doublejumps} (which is indeed model-dependent), but that can be clearly reabsorbed into the (fitting parameter) $\alpha$, giving a subleading correction to this. Naively, it can be understood as an expansion in combinatorial jump processes: as the distance between permutations becomes larger, their combinatory is described by higher polynomials in $k$, whose roots, except for the first ones, are model-dependent.

In the two models exhibited above we have, neglecting logarithmic corrections for RMPS
\begin{align}
     & \alpha_{\rm RMPS} = \frac{N}{\chi} \frac{d-1}{d} \Big(1 - \frac{d}{N(d-1)} +  \frac{1}{2}\beta_{\rm RMPS} \Big) ,\\&
      \beta_{\rm RMPS} = \frac{N}{\chi^2} \times {\rm const} ,
 \end{align}
and with the constant given in Eq.~\eqref{eq:alphabeta} depending on the types of gates considered. While for the random phase model (RPM) in the limit of large physical dimension we have, following from Eq.~\eqref{eq:RPM IPR},  
\begin{align}
     & \alpha_{\rm RPM} = \frac{N}{e^{\epsilon t}}   (1 +  2\beta_{\rm RMPS}), \\&
       \beta_{\rm RPM} = \frac{N}{e^{3\epsilon t/2}} .  
 \end{align}
 }
These equations imply therefore $\kappa_{\rm RMPS}=2$ (provided the substitution $\chi = 2^t$ is enforced), and  $\kappa_{\rm RPM}=3/2$.
\begin{figure*}[t!]
    \centering
    \includegraphics[width=\linewidth]{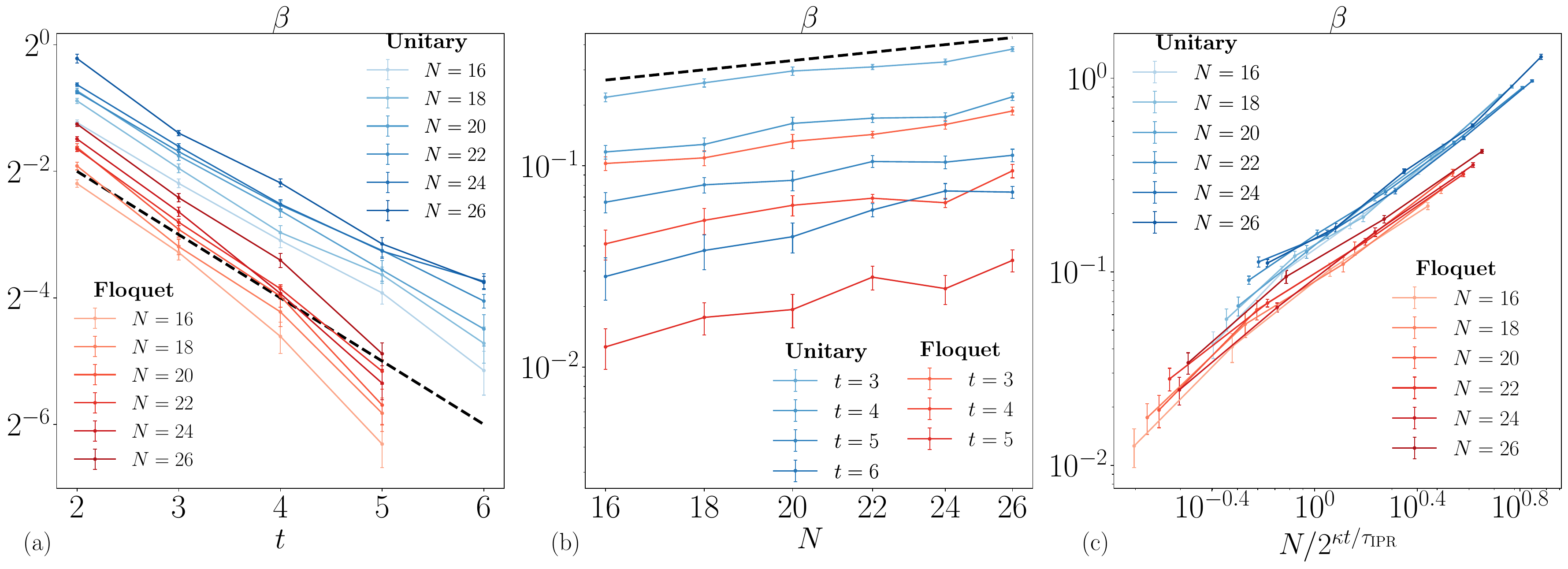}
    \caption{Panels \textbf{(a)},\textbf{(b)} and \textbf{(c)} show the behavior of $\beta$ with time and system size for the Floquet and unitary circuits. These values of $\beta$ have been obtained by fitting the distribution in Eq.~\eqref{eq:Haar conjecture distribution} to realizations of overlaps of the circuit (between 50k and 200k samples). The error bars indicate the standard deviation of the estimator. Panel \textbf{(a)} shows that $\beta$ decreases approximately as $2^{-t}$ (dashed line) at constant system size $N$. Panel \textbf{(b)} shows that $\beta$ increases as $N$ (dashed line) at constant time $t$. Panel \textbf{(c)} reveals that $\beta$ scales as $N/2^{\kappa t/\tau_{\mathrm{IPR}}}$, independently of the system size. We find $\kappa_U=2.74\pm 0.07$ and $\kappa_F= 3.13\pm 0.09$. (We separated the Floquet and Unitary data artificially by applying a factor of 1.5 to the Unitary data.) }
    \label{fig:plotsbeta}
\end{figure*}
We now benchmark our analytical predictions against extensive numerical simulations. 
We begin by focusing on brickwork unitary and orthogonal circuits for qubit systems, as described in Sec.~\ref{subsec:bwcircuit}, where odd and even layers alternate at each time step. In addition, we study a Floquet circuit in which each gate is fixed throughout the evolution. Specifically, we consider the Kicked Ising Model (KIM)~\cite{Tirrito2024}. Its Floquet operator is given by
\begin{equation}
    \label{eq:Floquet operator}
    U_\mathrm{F}= \exp\Bigl(-i\bigl(b\sum_j X_j + h\sum_j Z_j + J\sum_j Z_j Z_{j+1}\bigr)\Bigr),
\end{equation}
applied at each time step. Throughout this work, we set \(J=1\), \(b=(\sqrt{5}+5)/8\), and \(h=(\sqrt{5}+1)/4\). We do not expect our results to depend sensitively on these specific parameter choices, as long as the model remains non-integrable. 
To avoid basis-dependent effects, we initialize the KIM Floquet circuit in a random product state.

Our analysis starts with the time dependence of the parameter \(\alpha\), which we relate to the evolution of the state's purity. We then turn to the coefficient \(\beta\), and more generally, to the full joint distribution of \(\alpha\) and \(\beta\), whose moments are described by Eq.~\eqref{eq:conjec}.

\subsection{The scaling of {$\alpha$} and the evolution of the purity }

\begin{figure*}[t!]
    \centering
    \includegraphics[width=\linewidth]{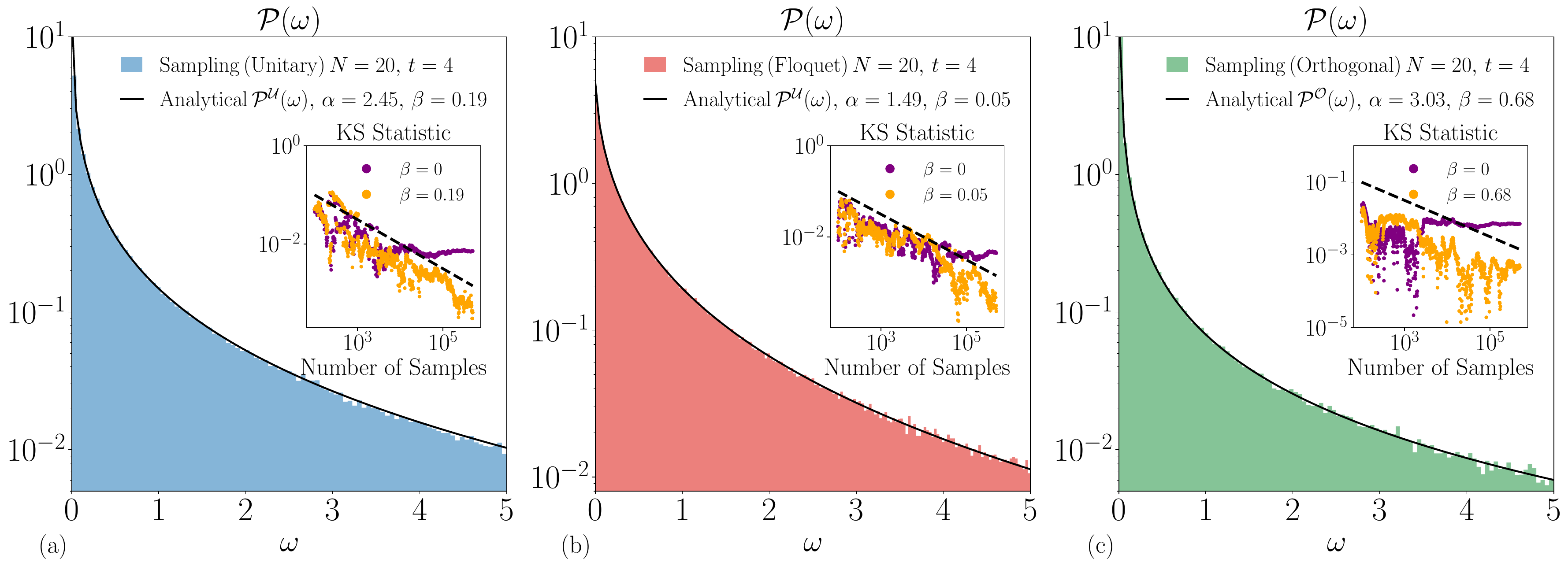}
    \caption{Panels \textbf{(a)},\textbf{(b)} and \textbf{(c)} show the sampled overlap distribution for $N=20$ and $t=4$ for the unitary, Floquet and orthogonal circuits respectively. We use $4.10 ^5$ samples and from them we fit the distribution (Eq.~\eqref{eq:Haar conjecture distribution}). Each inset highlights that considering finite-size corrections (non-zero $\beta$) allows us to reconstruct much better the overlap distribution than just ignoring them (zero $\beta$). In the first case, the KS statistic decreases as one over the square root of the number of samples (dashed line), indicating correctness of the distribution with finite $\beta$,  while the distribution with vanishing $\beta$ is detected to be incorrect.}
    \label{fig:plotsdistrib}
\end{figure*}
Our primary goal is to understand how the system approaches the Porter-Thomas distribution as the circuit depth and system size grow. 
To this end, we consider the deviation of the second participation entropy, $S_2(t,N)$, 
from its limiting value computed via the asymptotic Haar value. Specifically, we define
\begin{equation}
    \Delta S_2(t,N)=S_2(\infty,N)-S_2(t,N),
    \label{eq:delca}
\end{equation}
see Eq.~\eqref{eq:defiksk}. 
To further illustrate domain walls effects, we also consider the half-chain purity
\begin{equation}
    \mathcal{P}(t,N)=\mathrm{tr}\bigl(\rho_{N/2}^2\bigr),
\end{equation}
where \(\rho_{N/2}=\mathrm{tr}_{1,\dots,N/2}\bigl(|\Psi\rangle\langle\Psi|\bigr)\) is the reduced density matrix over half of the system. 
Both quantities are efficiently computable using the replica tensor network (RTN) approach with two replicas~\footnote{For Eq.~\eqref{eq:delca}, we consider the annealed average \(S_2 = -\log \mathbb{E}[I_2]\) since this quantity becomes self-averaging at moderate depths, as discussed in Ref.~\cite{turkeshi2024hilbert}. See also Refs.~\cite{nahum2017quantum,braccia_computing_2024,turkeshi2024quantummpembaeffectrandom,turkeshi2024magicspreadingrandomquantum} for details and~\cite{dataavail} for the code.}.
This allows us to uniquely determine the coefficient \(\alpha\) for large system sizes.

In Fig.~\ref{fig:RTNplots}(a), we compare the evolution of the purity \(\mathcal{P}(t,N)\) and \(\Delta S_2(t,N)\) for a qubit system of size \(N=128\). After a short transient, both quantities evolve at the same rate \(\alpha\). 
In this setting, we can analytically predict the timescale \(\tau_{\mathrm{IPR}}\) (associated with the inverse participation ratio) by examining the structure of the Weingarten matrix. 
Here, each layer of the brickwork circuit contributes dominantly to anticoncentration with a weight
\begin{equation}
    w_\mathcal{U} = \frac{2d}{d^2+1}.
\end{equation}
Focusing on two-replica calculations, and similarly to the RMPS case, we expand the circuit and obtain
\begin{equation}
    I^{\mathrm{BW},\mathcal{U}}_2 
    = I_2^\mathrm{Haar} \bigl(1+ c\, N\, w_\mathcal{U}^t + O(w_\mathcal{U}^{2t})\bigr).
\end{equation}
The subleading term is the dominant contribution to \(\Delta S_2(t,N)\), leading to the late-time scaling
\begin{equation}
    \Delta S_2(t,N)\sim \frac{N}{2^{\,t \lvert\log_2(w_\mathcal{U})\rvert}}.
\end{equation}
Hence, for qubit systems (\(d=2\)),
\begin{equation}
    \tau_{\mathrm{IPR},\mathcal{U}}
    = -\frac{1}{\log_d(w_\mathcal{U})}
    \approx 3.11.
\end{equation}
These observations align with the numerically extracted slope in Fig.~\ref{fig:RTNplots}(a). To further support these conclusions, in Fig.~\ref{fig:RTNplots}(b) we demonstrate a data collapse of \(\Delta S_2(t,N)\) for various \(N\), using the scaling variable \(N/2^{\,t/\tau_{\mathrm{IPR}}}\) and \(\tau_{\mathrm{IPR},\mathcal{U}}=3.11\). 
All system sizes and times coalesce onto a single curve, confirming our theoretical expectations.

For orthogonal circuits, the analysis is more intricate due to the absence of a single dominant contribution at early times. 
In Fig.~\ref{fig:RTNplots}(c), for \(N=128\), the transient period---before the purity and \(\Delta S_2(t,N)\) merge onto the same slope---is noticeably longer than in the unitary case [Fig.~\ref{fig:RTNplots}(a)]. 
Despite this, for \(t \gtrsim 16\), both quantities eventually align with the same timescale, yielding \(\tau_{\mathrm{IPR}}\approx \tau_{\mathrm{PUR}}\approx 3.2(1)\). 
Using this value in Fig.~\ref{fig:RTNplots}(d), we again observe a collapse of \(\Delta S_2(t,N)\) when plotted against \(N/2^{\,t/\tau_{\mathrm{IPR}}}\) with \(\tau_{\mathrm{IPR}}=3.2\). 
Overall, these observations highlight a central claim of this work: \textit{the purification and anticoncentration rates are closely related in chaotic quantum systems.}

\subsection{The $\alpha,\beta$ distribution}

We then supplement the RTN approach by analyzing the full distribution of overlaps with the computational basis in chaotic systems. This provides an unbiased estimate of \(\alpha\) and captures subsubleading corrections. We perform matrix product state (MPS) simulations using ITensor~\cite{itensor} for small depths without any truncation, and then fit the resulting overlap distribution with Eq.~\eqref{eq:Haar conjecture distribution} via a maximum likelihood estimation. 
Consistent with our previous arguments, \(\alpha\) should track \(\Delta S_2\) once finite-size and transient effects are negligible. 
These corrections are encoded in \(\beta\), which we anticipate to be subleading at late times; see also Fig.~\ref{fig:plotsbeta} below. 
Our results, for \(24\leq N\leq 42\) and sampling \(\mathcal{N} = 2 \times 10^4\) disorder realizations, are presented in Fig.~\ref{fig:plotsalpha}.

First, across all three models, \(\alpha\) and the system’s purity trace each other, as exemplified in Fig.~\ref{fig:plotsalpha}(a) for multiple system sizes. 
Focusing on each individual model, Fig.~\ref{fig:plotsalpha}(b) shows data collapse for the brickwork unitary circuit with \(\tau_{\mathrm{IPR}}=3.11\), consistent with the RTN estimate. 
In the orthogonal case, Fig.~\ref{fig:plotsalpha}(c), the best collapse occurs for \(\tau_{\mathrm{IPR}}\approx 3.32\), slightly larger than the RTN value (\(\sim 3.2\)). This minor discrepancy is expected, given that the timescale itself evolves significantly between early and late times for orthogonal circuits. Additionally, MPS simulations with individual disorder realizations are restricted to relatively short times.  Finally, for the KIM Floquet evolution, Fig.~\ref{fig:plotsalpha}(d) shows a data collapse with \(\tau_{\mathrm{IPR}} \approx 2.47\).

We now turn to the finite-size contributions controlled by \(\beta\). From Eq.~\eqref{eq:alphabeta}, we expect two main features: (i) at fixed time, \(\beta\) increases linearly with system size \(N\), and (ii) at fixed \(N\), \(\beta\) decays exponentially with time. 
To test this, we study the short-time regime of the unitary and Floquet circuits for \(N=16,18,\dots,26\), where large \(\beta\) values are expected. 
Because \(\beta\) is more challenging to pinpoint precisely (requiring a larger number of samples, here \(\mathcal{N}=2\times 10^5\)), the results are shown in Fig.~\ref{fig:plotsbeta}(a,b). 
In panel (a), we observe an exponential decay of \(\beta\) for each \(N\). In panel (b), at short times, \(\beta\propto N\). 
A sharper test is inspired by Conjecture~\ref{conjecture}, which posits
\begin{equation}
    \beta \sim {N}/{2^{\frac{\kappa t}{\tau}}}.
\end{equation}
We find \(\kappa_U = 2.74\) and \(\kappa_F = 3.13\) as best-fit values for our limited dataset. While numerics cannot decisively confirm this scaling, Fig.~\ref{fig:plotsbeta}(c) supports a qualitatively good agreement.

\begin{figure}[t]
    \centering
\end{figure}

Finally, we show how crucial it is to include the term \(\beta\) in reproducing overlap distributions, particularly at short times and relatively small \(N\), see Fig. \ref{fig:plotsdistrib}. There we compare the empirical distribution for a single instance of the unitary, Floquet and orthogonal circuits with \(N=20\) and \(t=4\) to the analytical form in Eq.~\eqref{eq:Haar conjecture distribution}, showing excellent agreement. 
Although \(\beta\) can be visually subtle in certain regimes, a Kolmogorov-Smirnov (KS) test quantitatively confirms its importance: 
the KS statistic \(\mathrm{KS}(F_\mathcal{N},F)=\sup_{\omega}\bigl|F_\mathcal{N}(\omega)-F(\omega)\bigr|\) decreases as \(1/\sqrt{\mathcal{N}}\) when \(\beta\) is fitted, but saturates if we set \(\beta=0\). 
As the insets of Fig.~\ref{fig:plotsbeta} show, including \(\beta\) substantially improves agreement with the empirical distribution.

\subsection{Applications to XEB benchmarking}
The knowledge of the IPR at finite times allows to benchmark the quantum circuit results even at small depths. Indeed, as explored in recent works \cite{arute2019quantum,PRXQuantum.5.010334,ware2023sharpphasetransitionlinear,Morvan2023}, an experimentally accessible proxy for the fidelity of a given quantum circuit is the Cross-Entropy Benchmarking
\begin{equation}
    \mathrm{XEB}=\Ex(\omega_{\mathrm{ideal}}\omega_{\mathrm{noisy}})-1,
\end{equation}
{i.e., measuring the correlation between the ideal overlap distribution of the bitstrings (typically computed in a classical machine) and the real one coming from a quantum computer. At large times or circuit depths, the XEB converges to the fidelity of the circuit, which in the limit of small $\varepsilon$ (weak noise) is known to be given by $F=(1- \varepsilon)^{\frac{Nt}{2}}$ where $1-\varepsilon$ is the fidelity of each single gate \cite{Dalzell2024,ware2023sharpphasetransitionlinear}. Given a unitary random circuit with, for example, a physically relevant global depolarizing noise in each gates (where with probability $\varepsilon$ each gate is replaced by a dephasing quantum channel), it is known that in the weak noise limit $\varepsilon N \to 0$, one can relate the XEB to the $I_2$ of the clean system, using a simple toy model for dephasing, where with probability $F$ the output of the circuit is the correct quantum state, and with probability $1-F$ is a classical string. This implies that 
\begin{equation}
    {\rm XEB} = F (D I_2(t) - 1)
\end{equation}
which clearly gives ${\rm XEB} \to F $ in the limit of large depth $t \gg \log N$ when $D I_2=2!$, providing this way a way to estimate the global fidelity $F$. Our results show that $F$ can instead be determined at finite, \textit{remarkably small values of depths $t \lesssim \log N$}, using the expression of $I^{\alpha,\beta}_2$ in terms of the (fitted) parameters $\alpha(t)$ and $\beta(t)$ that can be easily determined in a classical tensor network simulation. We report the results in Fig.~\ref{fig:plotxeb} where we show that while the normal estimation of fidelity is several orders of magnitudes off at short times, using our method one can already have a good predictions at very small depths,  allowing this way to benchmark quantum machines with much larger system's sizes using tensor network simulations. 

\begin{figure}[btbp]
    \centering
    \includegraphics[scale=1.0]{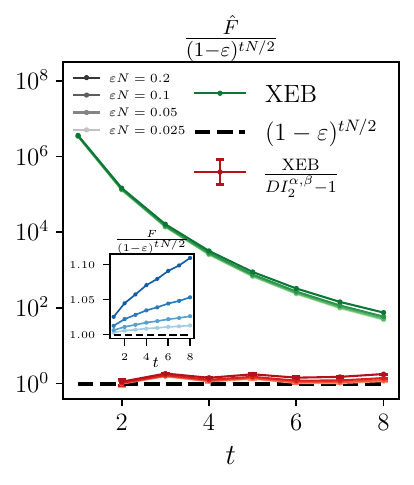}
    \caption{Global fidelity estimators $\hat{F}$,  for a noisy random unitary brickwork circuit with $N=64$, affected by depolarizing gate noise at local error rates of $\varepsilon N=0.025,\,0.05,\,0.1,\,0.2$ where darker shades indicate higher error rates. The standard XEB diverges from the circuit's fidelity, given by $(1- \varepsilon)^{tN/2}$ in the low noise limit (dashed black line),  at low depth. To improve fidelity estimation, we assume noise acts as a global depolarizing channel, leading to the estimator $\hat{F}={\mathrm{XEB}}/({D I^{\alpha,\beta}_2-1})$, with $I^{\alpha,\beta}_2$ defined in Eq.~\eqref{eq:conjec} and obtained from a noiseless tensor network simulation, and by fitting parameters $\alpha(t)$ and $\beta(t)$ at each time.  The inset illustrates how close the true fidelity $F$ is to its weak noise limit. All quantities are rescaled by $(1- \varepsilon)^{tN/2}$.}
    \label{fig:plotxeb}
\end{figure}

\section{Conclusion}
Although the microscopic details of quantum circuits affect their dynamics, we have shown that their anticoncentration properties are universal or they can be espressed in universal form, allwoing to obtain predictions that go well beyond the normal large time and large systems' sizes limits. Our theoretical framework, supported by large-scale numerical simulations, reveals that random tensor network states, random matrix product states, and brickwork circuits from various ensembles all display the same scaling behavior and universal form of the overlap distribution. 

A key insight is the universal crossover identified through RMPS, which governs both the leading and subleading corrections to Porter-Thomas statistics. This crossover is confirmed by data collapse in unitary and orthogonal circuits, further validated by extensive simulations of the Kicked Ising Model. 
Using Weingarten calculus and RTN methods, we characterized these finite-size corrections, showing that they depend only on a small set of parameters independent of the circuit architecture. The domain walls picture of anticoncentration provides an intuitive explanation for these corrections.

Our results set the stage for further investigations. As discussed in Ref.~\cite{Lami2025}, anticoncentration is closely tied to higher-order design properties and the frame potential, implying that similar scaling arguments should hold there, subject to a suitable rescaling of the characteristic time \(\tau\). Exploring higher-dimensional lattice models and implications for quantum complexity theory are natural next steps. 
From a practical perspective, such universal form could inform quantum algorithm design and error mitigation in near-term devices, by revealing the fundamental statistical constraints on random states in high-dimensional Hilbert spaces.

Overall, these findings underscore profound connections between quantum many-body dynamics, random matrix theory, and statistical physics. By elucidating the emergence of universal behavior in chaotic quantum systems, we anticipate broader implications for both fundamental physics and quantum technologies.

\begin{acknowledgments}
We thank Max McGinley, Romain Vasseur, P. Sierant, E. Tirrito, and Andrea De Luca for discussions and collaborations on related topics. 

B.M. and X.T. acknowledge DFG Collaborative Research Center (CRC) 183 Project No. 277101999 - project B01 and DFG under Germany's Excellence Strategy – Cluster of Excellence Matter and Light for Quantum Computing (ML4Q) EXC 2004/1 – 390534769. J.D.N., A.S., and G.L. are funded by the ERC Starting Grant 101042293 (HEPIQ) and the ANR-22-CPJ1-0021-01.

\textbf{Data availability.}
The numerical data for this work are given in Ref.~\cite{dataavail}, available at publication.

\end{acknowledgments}

\bibliography{biblio}
\end{document}